\documentclass[12pt]{article}
\usepackage{graphicx,psfrag,epsf,color}

\catcode`@=11
\newcount\@tempcntc
\def\@citex[#1]#2{\if@filesw\immediate\write\@auxout{\string\citation{#2}}\fi
  \@tempcnta\z@\@tempcntb\m@ne\def\@citea{}\@cite{\@for\@citeb:=#2\do
    {\@ifundefined
       {b@\@citeb}{\@citeo\@tempcntb\m@ne\@citea\def\@citea{,}{\bf ?}\@warning
       {Citation `\@citeb' on page \thepage \space undefined}}%
    {\setbox\z@\hbox{\global\@tempcntc0\csname b@\@citeb\endcsname\relax}%
     \ifnum\@tempcntc=\z@ \@citeo\@tempcntb\m@ne
       \@citea\def\@citea{,}\hbox{\csname b@\@citeb\endcsname}%
     \else
      \advance\@tempcntb\@ne
      \ifnum\@tempcntb=\@tempcntc
      \else\advance\@tempcntb\m@ne\@citeo
      \@tempcnta\@tempcntc\@tempcntb\@tempcntc\fi\fi}}\@citeo}{#1}}
\def\@citeo{\ifnum\@tempcnta>\@tempcntb\else\@citea\def\@citea{,}%
  \ifnum\@tempcnta=\@tempcntb\the\@tempcnta\else
   {\advance\@tempcnta\@ne\ifnum\@tempcnta=\@tempcntb \else \def\@citea{--}\fi
    \advance\@tempcnta\m@ne\the\@tempcnta\@citea\the\@tempcntb}\fi\fi}
\catcode`@=12

\begin{document}

\begin{titlepage}

\begin{flushright}
PITHA 08/27\\
SFB/CPP-08-83 \\
MPP-2008-132\\
October 14, 2008
\end{flushright}

\vskip1.5cm
\begin{center}
\Large\bf\boldmath
Higgs couplings in the MSSM at large $\tan\beta$
\end{center}

\vspace{0.7cm}
\begin{center}
{\sc M.~Beneke}$^{a,b}$, {\sc P. Ruiz-Femen\'\i a}$^a$, and 
{\sc M. Spinrath}$^{c}$\\[6mm]
{\small \it $^a$ Institut f{\"u}r Theoretische Physik~E,
RWTH Aachen University,\\
D--52056 Aachen, Germany}\\[0.2cm]
{\small $^b$ \it 
Institut f\"ur Theoretische Physik, 
Universit\"at Z\"urich,\\
CH -- 8057 Z\"urich, Switzerland}\\[0.2cm]
{\small \it $^c$  Max-Planck Institut f\"ur Physik,
F\"ohringer Ring~6,\\
D--80805 M\"unchen, Germany}
\end{center}

\vspace{1.5cm}
\begin{abstract}
\noindent
We consider $\tan\beta$-enhanced quantum effects in the minimal 
supersymmetric standard model (MSSM) including 
those from the Higgs sector. To this end, we match the MSSM to 
an effective two-Higgs doublet model (2HDM), assuming that all SUSY particles 
are heavy, and calculate the coefficients of the 
operators that vanish or are suppressed in the MSSM at tree-level. 
Our result clarifies the dependence of the large-$\tan\beta$ resummation 
on the renormalization convention for $\tan\beta$, and 
provides analytic expressions for the Yukawa and trilinear Higgs 
interactions.  The numerical effect is analyzed by means of a parameter scan, 
and we find that the Higgs-sector effects, where present, are typically 
larger than those from the ``wrong-Higgs'' Yukawa couplings  in the 
2HDM.
\end{abstract}
\end{titlepage}

\section{Introduction}

It is well-known that loop effects in the minimal supersymmetric 
extension of the standard model (MSSM) can become large when the ratio 
$\tan\beta = v_u/v_d$ of the two Higgs vacuum expectation values is 
sizeable. An important example 
is the Higgs coupling to bottom quarks, 
$H b\bar b$~\cite{Dabelstein:1995js,Coarasa:1995yg}. 
In the limit of heavy superpartner particle masses,  
the relevant couplings are \cite{Carena:1999py}
\begin{eqnarray}
&& y_D H_d\epsilon Q D -\delta\tilde y_D H_u^C\epsilon Q D +{\rm h.c.}
\nonumber\\
&& \to -(y_b v_d+\delta \tilde y_b v_u)\,\bar b_R b_L 
 - y_b H_d^0 \bar b_R b_L  - \delta \tilde y_b H_u^{0*}\bar b_R b_L 
+ {\rm h.c.}
\end{eqnarray}
This includes a coupling $\delta \tilde y_b$ to the ``wrong'' Higgs 
doublet, which is forbidden by supersymmetry at tree level. 
The quantity $\delta \tilde y_b$ is an ordinary one-loop effect 
and not enhanced in any way. However, due to the large 
ratio of vacuum expectation values (vevs) the bottom quark mass 
\begin{equation}
m_b = y_b v_d+\delta \tilde y_b v_u = 
m_b^{(0)}\left(1+\frac{\delta \tilde y_b}{y_b}\tan\beta\right)
\end{equation}
receives large $\tan\beta$-enhanced corrections to its 
tree-level value $m_b^{(0)}=y_b v_d$~\cite{Hall:1993gn}. So does 
the $H b\bar b$ coupling, when $y_b$ is eliminated in favour of $m_b$
as is usually done. In general, while the loop effect is 
not large by itself, it provides large corrections to couplings 
or observables, which are suppressed by the 
small vacuum expectation value $v_d$ or which vanish at tree level. 
This type of relative enhancement is clearly a one-loop effect, 
which does not repeat itself in higher loop orders. 

Here we investigate other effects of 
this type from the Higgs sector, where loop effects can lead to 
a mixing of the $H_u$ and $H_d^*$ fields through two- and four-point 
interactions, which indirectly modify the Yukawa couplings to 
the physical Higgs bosons. We also address the issue of renormalization 
conventions for $\tan\beta$ and their stability against radiative 
corrections, as investigated in a numerical approach in 
\cite{Freitas:2002um,Baro:2008bg}. 
The indirect effects from the Higgs sector have 
not been considered explicitly in the literature, although 
our results are implicit in one-loop calculations of MSSM Higgs 
decay rates~\cite{Dabelstein:1995js,Coarasa:1995yg}. However, 
these calculations are usually available only in numerical form.
Having simple analytic expressions at hand is often useful 
to acquire a better understanding of parameter dependences, 
especially for the MSSM with its large parameter space.
Higgs-sector effects from two-point interactions have 
been considered recently in~\cite{Freitas:2007dp}, while 
their consequences for $B\bar B$ mixing are discussed 
in~\cite{Trine:2007ma}.

The theoretical issues can be exposed most clearly in the 
decoupling limit, assuming that all superpartner particles are heavy, 
of order $M_{\rm SUSY}$, while the standard model (SM) particles 
together with all five physical Higgs bosons are light, 
at most $O(M_{\rm EW})$, the electroweak scale. In practice, the 
decoupling limit may be a good approximation for $M_{\rm SUSY}$ 
as small as a few hundred GeV. The assumed hierarchy 
implies a certain amount of fine-tuning, since the soft SUSY
breaking parameters and the $\mu$ parameter must be $O(M_{\rm SUSY})$, 
while $\hat m_{u,d}^2\equiv |\mu|^2+m_{u,d}^2$ are 
$O(M_{\rm EW})$ to make the Higgs bosons light. 
The Higgs-mixing parameter $b$ must even be  
much smaller than $M_{\rm EW}^2$ to achieve large $\tan \beta$. 
The first fine-tuning is the well-known ``little 
hierarchy problem'', here built in by hand by the assumption 
$M_{\rm SUSY}\gg M_{\rm EW}$, but the second is an extra fine-tuning 
of $b$ specific to the large-$\tan\beta$ scenario, which has 
received less attention up to now. To simplify 
the discussion we consider only the third family of fermions 
and assume CP conservation in the MSSM. Including 
the CKM matrix and CP-violating MSSM phases would be 
straightforward. Under these assumptions we match the MSSM 
to an effective general two-Higgs doublet model (2HDM), and 
determine the physical Higgs fields, the Yukawa couplings, 
and trilinear Higgs couplings from the one-loop corrected 
effective 2HDM.

\section{Matching the MSSM to the 2HDM}

The Higgs and Yukawa terms of the MSSM Lagrangian are given by
\begin{eqnarray}
{\cal L} &=& (D_\mu H_d)^\dagger (D^\mu H_d) 
+ (D_\mu H_u)^\dagger (D^\mu H_u) -\hat m_d^2 H_d^\dagger H_d 
-\hat m_u^2 H_u^\dagger H_u 
\nonumber\\
&&
- \,b\,(H_d\epsilon H_u - H_u^\dagger \epsilon H_d^\dagger)
- \frac{g_2^2}{2} \, (H_d^\dagger H_u) (H_u^\dagger H_d) 
-\frac{1}{8} (g_2^2+g_1^2) \,(H_d^\dagger H_d-H_u^\dagger H_u)^2
\nonumber\\
&&  - \,y_t H_u\epsilon Q U +y_b H_d\epsilon Q D
+y_\tau H_d\epsilon L E+{\rm h.c.}
\label{mssm}
\end{eqnarray}
The conventions are as follows: the antisymmetric $2\times 2$ matrix 
$\epsilon$ is defined with $\epsilon_{12}=-1$. The hypercharge 
U(1)$_{\rm Y}$ and SU(2) couplings are $g_1$ and $g_2$, respectively. 
The quark fields are $Q_\alpha=(t_{L\alpha},b_{L\alpha})^T$,
$U^\alpha=(t_R^*)^\alpha$, $D^\alpha=(b_R^*)^\alpha$, the leptons 
 $L_\alpha=(\nu_{L\alpha},\tau_{L\alpha})^T$,
$E^\alpha=(\tau_R^*)^\alpha$, where $\alpha=1,2$ denotes a Weyl 
spinor index.
For the Higgs doublets we also use the notation $\Phi_1=H_d$, 
$\Phi_2=H_u^C=\epsilon H_u^*$, such that $\Phi_1$ and $\Phi_2$ 
both have hypercharge $-1/2$. Our MSSM conventions follow 
\cite{Rosiek:1989rs} except for $y_b = - [y_b]_{\rm Rosiek}$,
$y_\tau = - [y_\tau]_{\rm Rosiek}$, $b= -[m_{12}^2]_{\rm Rosiek}$ and
$A_t= - [A_t]_{\rm Rosiek}$. In this way our definitions 
of the Yukawa couplings and $A$ parameters are consistent
with the SPA convention~\cite{AguilarSaavedra:2005pw}.

\subsection{Parameters and renormalization of the MSSM}

The relevant parameters of the MSSM Higgs sector are $g_1$, $g_2$, 
$\hat m^2_{u,d}$, $b$, $y_{b,t}$ and the two vevs $v_{u,d}$. It appears 
to be contrary to the spirit of interpreting the 
MSSM as the high-scale theory and the 2HDM as the low-energy 
effective theory to introduce the vevs as MSSM parameters, since 
the vevs should be found {\em a posteriori} by minimizing the 2HDM 
effective potential. However, since we wish to use standard 
MSSM renormalization conventions, and moreover introduce 
$\tan\beta$ as one of the MSSM parameters, we substitute 
\begin{eqnarray}
H_d \to \left(\begin{array}{c} v_d+H_d^0\\
H_d^-\end{array}\right),
\qquad
H_u \to \left(\begin{array}{c} H_u^+ \\ v_u+H_u^0
\end{array}\right)
\label{introducevevs}
\end{eqnarray}
in the Lagrangian (\ref{mssm}). In this way, after further 
fixing the renormalization convention, the parameter 
$\tan\beta = v_u/v_d$ is well-defined, contrary to the general 
2HDM, where $\tan\beta$ is unphysical, since it can assume any value 
by performing a rotation in the $(\Phi_1,\Phi_2)$ fields. In the MSSM 
such rotations are excluded by holomorphy of the 
superpotential.
 
An equivalent set of parameters consists of 
\begin{equation}
e,M_W, M_Z, M_A,\tan\beta,t_{u,d}, m_{b,t},
\end{equation} 
related to the previous set by
\begin{eqnarray}
e &=& \frac{g_1 g_2}{\sqrt{g_1^2+g_2^2}},
\nonumber\\
M_W&=& \frac{g_2}{\sqrt{2}}\, v,\quad
M_Z = \frac{1}{\sqrt{2}} \sqrt{g_1^2+g_2^2} \,v,
\nonumber\\
M_A^2&=&\hat m_u^2+\hat m_d^2,\quad
\tan\beta =\frac{v_u}{v_d},
\nonumber\\
t_d&=& \hat m_d^2 v_d+\frac{1}{4}\,(g_1^2+g_2^2) \,v_d \,(v_d^2-v_u^2)-b v_u,
\nonumber\\
t_u&=& \hat m_u^2 v_u+\frac{1}{4}\,(g_1^2+g_2^2) \,v_u \,(v_u^2-v_d^2)-b v_d,
\label{parameterrel}
\end{eqnarray}
with $v=\sqrt{v_u^2+v_d^2}$. The quark masses are given by 
$m_{b,t} = y_{b,t} v_{d,u}$ at tree level.
For vanishing tadpole couplings 
$t_{u,d}$, $M_A$ is the mass of the CP-odd neutral Higgs boson. 

The parameters so far are bare parameters. We introduce renormalized 
parameters by substituting $p\to p_0\to p+\Delta p$, where now $p_0$ 
($p$) is the bare (renormalized) parameter and $\Delta p$ the corresponding 
counterterm. The renormalization conventions are specified by 
imposing conditions on the parameters of the second set: 
$e,M_W,M_Z, M_A$ are defined by the conventional on-shell 
renormalization conditions for the electric charge and the particle 
pole masses, the tadpole couplings $t_{u,d}$ by the condition that 
the tadpoles (one-point functions) vanish. The renormalization 
condition for $\tan\beta$ and the Yukawa couplings, 
and  the definition of $m_{b,t}$ beyond tree 
level will be discussed later. 
The counterterms $\Delta p$ for the parameters of the first set 
are then defined by requiring that (\ref{parameterrel}) holds for 
the bare as well as the renormalized parameters. Specifically, 
for $\tan\beta$ this implies
\begin{equation}
\tan\beta +\Delta t_\beta = [\tan \beta]_0 = 
\frac{[v_u]_0}{[v_d]_0} 
= \frac{v_u+\Delta v_u}{v_d+\Delta v_d} 
= \frac{v_u}{v_d} +  \frac{v_u}{v_d} \left(\frac{\Delta v_u}{v_u} -
\frac{\Delta v_d}{v_d}\right) + \ldots.
\end{equation}
Employing $\tan\beta = v_u/v_d$, the one-loop relation 
\begin{equation}
\Delta t_\beta = \tan\beta \left(\frac{\Delta v_u}{v_u} -
\frac{\Delta v_d}{v_d}\right)
\label{tanbcounter}
\end{equation}
follows. Similarly, one obtains a set of equations relating the shifts of 
all parameters of the second set to those of the first.

We do not need to discuss the renormalization of
superpartner masses and the $A$ and $\mu$ terms, since these 
occur only in loops. In a 
one-loop calculation, it is sufficient to know these parameters 
at tree-level. Other SM parameters like the strong coupling or 
CKM matrix play no role here. For the subsequent discussion of 
$\tan\beta$-enhanced terms field renormalization in the MSSM 
can also be ignored. Thus, we shall not introduce renormalization 
factors $Z_{u,d}$ for the SU(2)-doublet fields. The divergent 
parts of the vev counterterms $\Delta v_{u,d}$ are identical 
to those in $Z_{u,d}$, since the theory in the broken phase can 
be rendered finite by the same counterterms as the unbroken 
theory before the shift (\ref{introducevevs}). Thus, when we 
ignore field renormalization, we should formally consider 
$\Delta v_{u,d}$ and consequently $\Delta t_\beta$ as finite 
quantities, which indeed they are in the large-$\tan \beta$ 
approximation, whose values are nevertheless fixed by the imposed 
renormalization conventions.

\subsection{The effective 2HDM Lagrangian} 

We integrate out the heavy SUSY particles and expand the MSSM 
effective action in local operators. The
result is a general 2HDM, where the Higgs-Yukawa terms read
\begin{eqnarray}
{\cal L}_{\rm 2HDM} &=& Z_{ab} (D_\mu\Phi_a)^\dagger (D^\mu\Phi_b) - 
m_{ab}^2 \Phi_a^\dagger \Phi_b
\nonumber\\
&&-\,\frac{\lambda_1}{2} (\Phi_1^\dagger \Phi_1)^2 
-\frac{\lambda_2}{2} (\Phi_2^\dagger \Phi_2)^2
-\lambda_3 (\Phi_1^\dagger \Phi_1)\,(\Phi_2^\dagger \Phi_2)
-\lambda_4 (\Phi_1^\dagger \Phi_2)\,(\Phi_2^\dagger \Phi_1)
\nonumber\\
&&-\,\left[ \frac{\lambda_5}{2} (\Phi_1^\dagger \Phi_2)^2 
+ \Big(\lambda_6 \Phi_1^\dagger \Phi_1+\lambda_7 \Phi_2^\dagger
\Phi_2\Big)\,\Phi_1^\dagger \Phi_2 + {\rm h.c.}\right] 
\label{twohdm}\\
&&+Y_b \Phi_1\epsilon Q b + Y_t \Phi_2^\dagger Q t 
+Y_\tau \Phi_1\epsilon L \tau - \delta \tilde y_b \Phi_2\epsilon Q b 
+\delta \tilde y_t \Phi_1^\dagger Q t - \delta \tilde y_\tau 
\Phi_2\epsilon L \tau +{\rm h.c.}\quad 
\nonumber
\end{eqnarray}
neglecting operators with dimension higher than four.
The last line equals 
\begin{eqnarray*}
&&Y_b H_d\epsilon Q b - Y_t H_u\epsilon Q t +Y_\tau H_d\epsilon L \tau
- \delta \tilde y_b H_u^C\epsilon Q b + \delta \tilde y_t
H_d^C\epsilon Q t - \delta \tilde y_\tau H_u^C\epsilon L \tau +{\rm h.c.} 
\end{eqnarray*}
in terms of the $H_{u,d}$ fields, so the Yukawa couplings take their 
standard form. Since we assumed CP conservation in the MSSM, the 
matrices $Z_{ab}$, $m_{ab}^2$ ($a,b=1,2$) and the couplings 
$\lambda_i$ are real. Note that the kinetic terms of the Higgs 
fields are not yet canonical, because we match 1PI Green functions. 
The kinetic terms of the fermions can be disregarded, since their 
renormalization is an ordinary one-loop correction to an unsuppressed 
tree term. At tree-level
\begin{eqnarray}
&& Z_{ab} =\delta_{ab}, 
\nonumber\\
&& m_{11}^2 = \hat m_d^2, \quad
m_{22}^2 = \hat m_u^2,  \quad 
m_{12}^2 = m_{21}^2 = b,
\nonumber\\ 
&&\lambda_1=\lambda_2=\frac{1}{4} (g_1^2+g_2^{2}),  \quad
\lambda_3=\frac{1}{4} (g_2^2-g_1^{2}),  \quad
\lambda_4=-\frac{1}{2}g_2^2,
\nonumber\\ 
&& \lambda_5=\lambda_6=\lambda_7=0,
\nonumber\\ 
&& Y_{b,t,\tau} =y_{b,t,\tau},\quad 
\delta \tilde y_b = \delta \tilde y_t=\delta \tilde y_\tau=0. 
\end{eqnarray}
In the following the one-loop corrections to the couplings that vanish
or are small at tree-level are of particular interest: 
$\delta Z_{12}\equiv z_{12}$, $\delta m_{12}^2$, 
$\delta\lambda_{5,6,7}$, and $\delta \tilde y_{b,t,\tau}$. The result 
of the calculation is summarized in the next section.

\subsection{One-loop results}

\begin{figure}[t]
  \begin{center}
  \vspace*{-0.1cm}
  \includegraphics[width=0.9\textwidth]{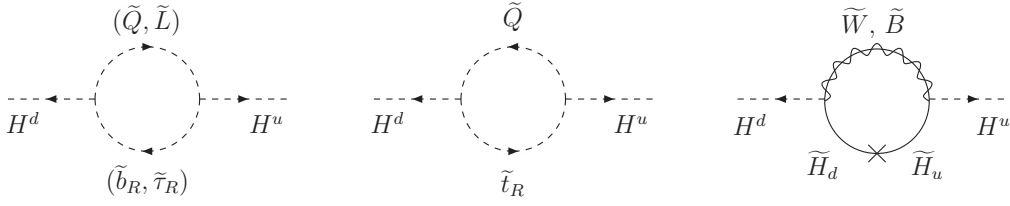}
  \vspace*{0cm}
  \caption{Diagrams entering the matching of $m_{12}^2$ and $z_{12}$.}
  \label{fig1}
  \end{center}
\end{figure}

All matching calculations are performed in the SU(2)$\times$U(1)$_Y$ 
symmetric ``phase'' of the MSSM.
The couplings $z_{12}$ and $m_{12}^2$ follow from an expansion of 
the off-diagonal $H_u H_d$ two-point function in its external momentum. 
The relevant one-loop diagrams with heavy superpartner particles in the loop 
are shown in Figure~\ref{fig1}. For the 
kinetic-mixing term we find
\begin{eqnarray}
z_{12} &=& \mu^*\,\Bigg\{
3 y_t A_t^*\,H_2\!\left(m_{\tilde t_R},m_{\tilde Q}\right) 
+ 3 y_b A_b^*\,H_2\!\left(m_{\tilde b_R},m_{\tilde Q}\right) 
+ y_\tau A_\tau^*\,H_2\!\left(m_{\tilde \tau_R},m_{\tilde L}\right) 
\nonumber\\
&& \hspace*{1.1cm}
+ \,3 g_2^2 M_2\,H_2\!\left(|\mu|,M_2\right) 
+ g_1^2 M_1\,H_2\!\left(|\mu|,M_1\right) 
\Bigg\}
\label{z121loop}
\end{eqnarray}
For the off-diagonal mass term, we write $m_{12}^2 = 
b+\delta b+[\Delta b]_{\rm heavy}$ and obtain for $\delta b$ 
the expression (\ref{z121loop}) with the replacement 
$ H_2(m_1,m_2) \to  - I_2(m_1,m_2)$, where 
\begin{eqnarray}
I_2(m_1,m_2) &=&\frac{1}{16\pi^2} \left(\frac{1}{\epsilon} 
-\ln\frac{m_1^2}{\nu^2}+i_2(x)\right) \,, 
\nonumber \\
i_2(x) &=& 1+\frac{x \ln x}{1-x} \,,\\
H_2(m_1,m_2) &=&\frac{1}{16\pi^2}\,\frac{1}{2 m_1^2}\,
h_2(x) \,,
\nonumber\\
h_2(x) &=& \frac{1-x^2+2 x \ln x}{(1-x)^3} \,,
\end{eqnarray}
and $x\equiv m_2^2/m_1^2$. The kinetic-mixing correction $z_{12}$ 
is finite, as it should be, while the correction to $m_{12}^2$ 
is ultraviolet divergent and requires regularization (here dimensional 
regularization, space-time dimension $d=4-2\epsilon$ and renormalization scale $\nu$). 
The $1/\epsilon$ pole is cancelled by the counterterm contribution 
$[\Delta b]_{\rm heavy}$ to $m_{12}^2$. 
The counterterm contribution is the difference between 
the MSSM and the 2HDM counterterm, which survives the matching 
relation, and cancels the singularity in the heavy-particle loops. 
In the following we shall drop the label ``heavy'' in the 
counterterm contributions. We remark that we have kept complex conjugation 
on the $\mu$ and $A$ parameters, where required, if they were complex. 
However, by our assumption of CP-conservation, $\mu$ and $A_{t,b,\tau}$
are real. Here and below the masses that enter the arguments of the loop 
functions are the gaugino and sfermion mass parameters in the 
soft supersymmetry-breaking part of the MSSM Lagrangian and not 
the physical mass parameters. The physical masses receive additional 
contributions at tree level
proportional to the vevs $v_{u,d}$ from electroweak symmetry breaking, 
which are of higher order in the $M_{\rm EW}/M_{\rm SUSY}$ expansion.

\begin{figure}[t]
  \begin{center}
  \vspace*{-0.1cm}
  \includegraphics[width=0.95\textwidth]{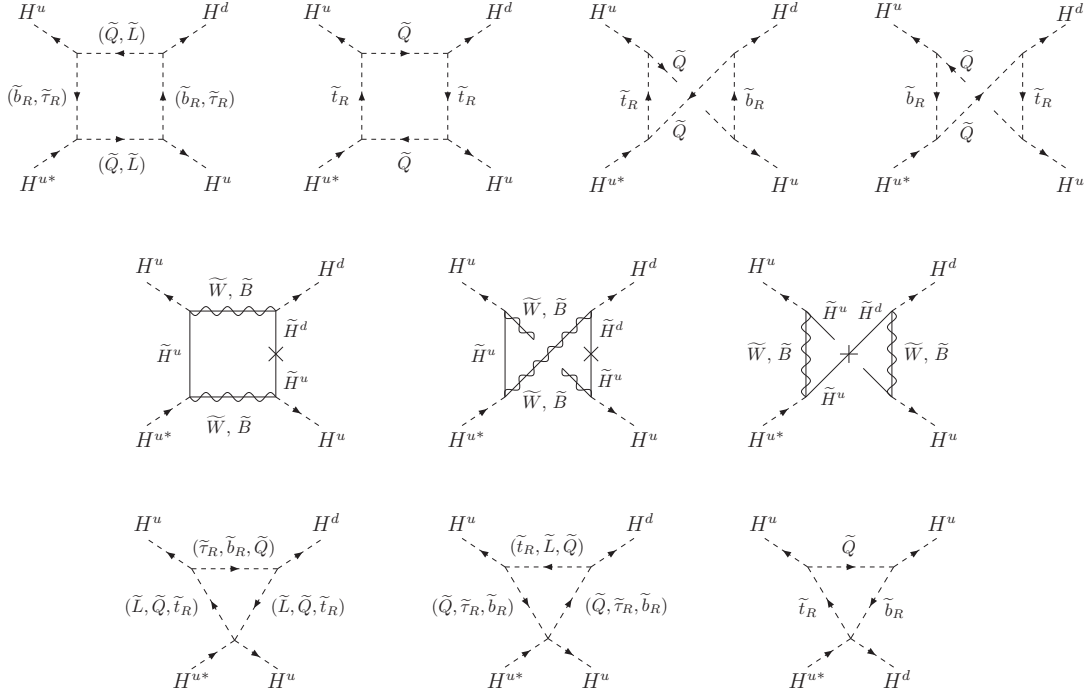}
  \vspace*{0.2cm}
  \caption{Diagrams entering the matching of $\delta\lambda_7$.}
  \label{fig2}
  \end{center}
\end{figure}

For the four-point Higgs couplings $\lambda_{5,6,7}$, which vanish
at tree level, we obtain the ultraviolet-finite one-loop
expressions
\begin{eqnarray}
\delta \lambda_5 &=& - \, 3(\mu^* A_t^*)^2 y_t^2 K_2(m_{\tilde Q},m_{\tilde t_R})
- 3(\mu^* A_b^*)^2 y_b^2 K_2(m_{\tilde Q},m_{\tilde b_R})
\nonumber\\[0.05cm]
&& - \, (\mu^* A_\tau^*)^2 y_\tau^2 K_2(m_{\tilde L},m_{\tilde \tau_R})
\nonumber\\[0.15cm] 
&& +\, (\mu^*)^2 M_1^2 \, g_1^4 \, K_2(|\mu|,M_1) 
+\, 3 (\mu^*)^2 M_2^2 \, g_2^4 \, K_2(|\mu|,M_2) 
\nonumber\\[0.1cm] 
&& +\,2 (\mu^*)^2 M_1 M_2 \, g_1^2 g_2^2 \, K_3(|\mu|,M_1,M_2) \,,
\\[0.4cm]
\delta \lambda_6 &=& - \, 3\mu^* A_b^* \,y_b^3\left\{
J_2(m_{\tilde Q},m_{\tilde b_R}) +
J_2(m_{\tilde b_R},m_{\tilde Q}) \right\} 
\nonumber\\ 
&& - \, \mu^* A_\tau^* \,y_\tau^3\left\{
J_2(m_{\tilde L},m_{\tilde \tau_R}) +
J_2(m_{\tilde \tau_R},m_{\tilde L}) \right\} 
\nonumber\\ 
&& +\,3\mu^* y_t A_t^*  \left(-\frac{1}{4}\right) \left\{
\left[g_2^2-\frac{g_1^2}{3}\right]
J_2(m_{\tilde Q},m_{\tilde t_R}) +
\left[\frac{4g_1^2}{3}\right]
J_2(m_{\tilde t_R},m_{\tilde Q}) \right\} 
\nonumber\\ 
&& +\,3\mu^* y_b A_b^* \left(-\frac{1}{4}\right) \left\{
\left[-g_2^2-\frac{g_1^2}{3}\right]
J_2(m_{\tilde Q},m_{\tilde b_R}) +
\left[-\frac{2 g_1^2}{3}\right]
J_2(m_{\tilde b_R},m_{\tilde Q}) \right\} 
\nonumber\\ 
&& +\,\mu^* y_\tau A_\tau^* \left(-\frac{1}{4}\right) \left\{
\left[-g_2^2+g_1^2\right]
J_2(m_{\tilde L},m_{\tilde \tau_R}) +
\left[-2 g_1^2\right]
J_2(m_{\tilde \tau_R},m_{\tilde L}) \right\}  
\nonumber\\[0.1cm] 
&& -\, 3 \mu^* \,|\mu|^2 A_t^* y_t^3 K_2(m_{\tilde Q},m_{\tilde t_R})
-  3 \mu^* \,|A_b|^2 A_b^* y_b K_2(m_{\tilde Q},m_{\tilde b_R})
\nonumber\\[0.2cm]
&& - \, \mu^* \,|A_\tau|^2 A_\tau^* y_\tau 
K_2(m_{\tilde L},m_{\tilde \tau_R})
\nonumber\\[0.2cm] 
&& -\, 3 \mu^* \,M_2 g_2^4 \,L_2(|\mu|,M_2) 
- \mu^* \,M_1 g_1^4 \,L_2(|\mu|,M_1) 
\nonumber\\[0.2cm] 
&& - \,\mu^* \,(M_2+M_1) \,g_2^2 g_1^2 \,L_3(|\mu|,M_2,M_1) \,,
\\[0.4cm]
\delta \lambda_7 &=& -\, 3\mu^* y_t A_t^* \,y_t^2\left\{
J_2(m_{\tilde Q},m_{\tilde t_R}) +
J_2(m_{\tilde t_R},m_{\tilde Q}) \right\} 
\nonumber\\ 
&& -\,3\mu^* y_t A_t^*  \left(-\frac{1}{4}\right) \left\{
\left[g_2^2-\frac{g_1^2}{3}\right]
J_2(m_{\tilde Q},m_{\tilde t_R}) +
\left[\frac{4g_1^2}{3}\right]
J_2(m_{\tilde t_R},m_{\tilde Q}) \right\} 
\nonumber\\ 
&& -\,3\mu^* y_b A_b^* \left(-\frac{1}{4}\right) \left\{
\left[-g_2^2-\frac{g_1^2}{3}\right]
J_2(m_{\tilde Q},m_{\tilde b_R}) +
\left[-\frac{2 g_1^2}{3}\right]
J_2(m_{\tilde b_R},m_{\tilde Q}) \right\} 
\nonumber\\ 
&& -\,\mu^* y_\tau A_\tau^* \left(-\frac{1}{4}\right) \left\{
\left[-g_2^2+g_1^2\right]
J_2(m_{\tilde L},m_{\tilde \tau_R}) +
\left[-2 g_1^2\right]
J_2(m_{\tilde \tau_R},m_{\tilde L}) \right\}  
\nonumber\\[0.1cm] 
&& -\, 3 \mu^* \,|A_t|^2 A_t^* y_t K_2(m_{\tilde Q},m_{\tilde t_R})
-  3 \mu^* \,|y_b \mu^*|^2 A_b^* y_b K_2(m_{\tilde Q},m_{\tilde b_R})
\nonumber\\[0.2cm]
&& - \, \mu^* \,|y_\tau \mu^*|^2 A_\tau^* y_\tau 
K_2(m_{\tilde L},m_{\tilde \tau_R})
\nonumber\\[0.2cm] 
&& -\, 3 \mu^* \,M_2 g_2^4 \,L_2(|\mu|,M_2) 
- \mu^* \,M_1 g_1^4 \,L_2(|\mu|,M_1) 
\nonumber\\[0.2cm] 
&& - \,\mu^* \,(M_2+M_1) \,g_2^2 g_1^2 \,L_3(|\mu|,M_2,M_1) \,.
\label{eq:lambda7}
\end{eqnarray}
The diagrams relevant for the computation of $\delta \lambda_7$ are shown 
in Figure~\ref{fig2}. The loop functions read
\begin{eqnarray}
J_2(m_1,m_2) &=&\frac{1}{16\pi^2} \left(-\frac{1}{m_1^2}\right) 
j_2(x) \,,
\nonumber\\
j_2(x) &=& \frac{1}{1-x}+\frac{x \ln x}{(1-x)^2} \,,\\
K_2(m_1,m_2) &=&\frac{1}{16\pi^2} \left(-\frac{1}{m_1^4}\right) 
k_2(x) \,,
\nonumber\\
k_2(x) &=& \frac{2}{(1-x)^2}+\frac{(1+x) \ln x}{(1-x)^3} \,,\\
K_3(m_1,m_2,m_3) &=&\frac{1}{16\pi^2} \left(-\frac{1}{m_1^4}\right) 
k_3(x,y) \,,
\nonumber\\
k_3(x,y) &=& \frac{1}{(1-x) (1-y)}+\frac{x\ln x}{(x-y) (1-x)^2}+
\frac{y\ln y}{(y-x) (1-y)^2} \,,
\nonumber\\
L_2(m_1,m_2) &=&\frac{1}{16\pi^2} \left(-\frac{1}{m_1^2}\right) 
l_2(x) \,,
\nonumber\\
l_2(x) &=& \frac{1+x}{(1-x)^2}+\frac{2 x\ln x}{(1-x)^3} \,,\\
L_3(m_1,m_2,m_3) &=&\frac{1}{16\pi^2} \left(-\frac{1}{m_1^2}\right) 
l_3(x,y) \,,
\nonumber\\
l_3(x,y) &=& \frac{1}{(1-x) (1-y)}+\frac{x^2\ln x}{(x-y) (1-x)^2}+
\frac{y^2\ln y}{(y-x) (1-y)^2} \,,
\end{eqnarray}
with $x\equiv m_2^2/m_1^2$, $y\equiv m_3^2/m_1^2$.

The loop-induced Yukawa couplings to the ``wrong'' Higgs field are 
given by 
\begin{eqnarray}
\delta \tilde y_{b} &=& 
- \, \frac{8}{3} \mu^* y_b M_3 \, g_3^2 J_3(M_3,m_{\tilde Q},m_{\tilde b_R})
- \, \mu^* y_b y_t A_t^* J_3(|\mu|,m_{\tilde Q},m_{\tilde t_R}) 
\nonumber\\
&& 
+ \, \frac{3}{2} \mu^* y_b M_2 \, g_2^2 J_3(|\mu|,M_2,m_{\tilde Q}) 
+ \, \frac{1}{9} \mu^* y_b M_1 \, g_1^2 J_3(M_1,m_{\tilde Q},m_{\tilde b_R}) 
\nonumber\\
&& + \, \frac{1}{3} \mu^* y_b M_1 \, g_1^2 J_3(|\mu|,M_1,m_{\tilde b_R}) 
+ \, \frac{1}{6} \mu^* y_b M_1 \, g_1^2 J_3(|\mu|,M_1,m_{\tilde Q})\,, 
\label{eq:Deltayb}
\\[0.2cm]
\delta \tilde y_{t} &=& 
\frac{8}{3} \mu^* y_t M_3 \, g_3^2 J_3(M_3,m_{\tilde Q},m_{\tilde t_R})
+ \, \mu^* y_b y_t A_b^* J_3(|\mu|,m_{\tilde Q},m_{\tilde b_R}) 
\nonumber\\
&& 
- \, \frac{3}{2} \mu^* y_t M_2 \, g_2^2 J_3(|\mu|,M_2,m_{\tilde Q}) 
+ \, \frac{2}{9} \mu^* y_t M_1 \, g_1^2 J_3(M_1,m_{\tilde Q},m_{\tilde t_R}) 
\nonumber\\
&& - \, \frac{2}{3} \mu^* y_t M_1 \, g_1^2 J_3(|\mu|,M_1,m_{\tilde t_R}) 
+ \, \frac{1}{6} \mu^* y_t M_1 \, g_1^2 J_3(|\mu|,M_1,m_{\tilde Q}) \,,
\label{eq:Deltayt}
\\[0.2cm]
\delta \tilde y_{\tau} &=& 
 \frac{3}{2} \mu^* y_\tau M_2 \, g_2^2 J_3(|\mu|,M_2,m_{\tilde L}) 
- \, \mu^* y_\tau M_1 \, g_1^2 J_3(M_1,m_{\tilde L},m_{\tilde \tau_R}) 
\nonumber\\
&&+ \, \mu^* y_\tau M_1 \, g_1^2 J_3(|\mu|,M_1,m_{\tilde \tau_R}) 
- \, \frac{1}{2} \mu^* y_\tau M_1 \, g_1^2 J_3(|\mu|,M_1,m_{\tilde L}) 
\,,
\label{eq:Deltaytau}
\end{eqnarray}
where
\begin{eqnarray}
J_3(m_1,m_2,m_3) &=&\frac{1}{16\pi^2} \left(\frac{1}{m_1^2}\right) 
j_3(x,y) \,,
\nonumber\\
j_3(x,y) &=& \frac{x\ln x}{(x-y) (1-x)}+
\frac{y\ln y}{(y-x) (1-y)} \,.
\end{eqnarray}
The expression for $\delta \tilde y_{b}$ agrees with the well-known 
result for the $\tan\beta$-enhanced correction to the bottom 
quark mass \cite{Carena:1999py,Hall:1993gn}, 
when the bino contribution and terms of order $M_{\rm EW}/
M_{\rm SUSY}$ are neglected. Our full results for $\delta \tilde y_{b}/y_b$
and $\delta \tilde y_{t}/y_t$ agree with the expressions for the 
correspondent quantities 
$\epsilon_0+\epsilon_Y y_t^2$ and 
$-[\epsilon^\prime_0+\epsilon_Y^\prime y_b^2]$ given  
in~\cite{Freitas:2007dp}, which were written in terms of the 
rescaled $A$-parameters, $A_{b,t}\to y_{b,t}A_{b,t}$.
Our result for $\delta \lambda_5$ agrees with~\cite{Trine:2007ma}, where the 
bino contributions have been neglected.

It is no surprise that the one-loop effects of interest are all 
proportional to the $\mu$-parameter, since the off-diagonal two-point 
terms and the $\lambda_{5,6,7}$ couplings are protected by a 
U(1)$_{\rm PQ}$ symmetry that is broken in the MSSM only by the 
$\mu$ and $b$ term \cite{Hall:1993gn}. Note that the quantum correction to 
$m_{12}^2$ is of $O(M_{\rm SUSY}^2)$, while the $b$ (or rather 
$m_{12}^2$) parameter 
was assumed to be of order $M_{\rm EW}^2/\tan\beta$. Thus, 
the large-$\tan\beta$ scenario requires more severe fine-tuning 
(by a factor of $\tan\beta$) than the one to maintain the 
``little hierarchy'' $M_{\rm EW}\ll M_{\rm SUSY}$. It therefore 
appears less natural in this respect than the generic MSSM.

\section{Higgs couplings}

We are now ready to derive the one-loop corrected Higgs couplings, 
which are $\tan\beta$-enhanced relative to their tree-level 
expressions. For any quantity ${\cal G}$  
we can write its NLO MSSM expression as 
\begin{equation}
{\cal G} = {\cal G}_{\rm tree} + {\cal G}_{\rm heavy} + 
{\cal G}_{\rm light},
\end{equation}
where ${\cal G}_{\rm heavy}$ is the contribution from  
the scale $M_{\rm SUSY}$, in practice from
loops containing SUSY particles including the corresponding 
counterterms, and ${\cal G}_{\rm light}$ the remaining 
contribution from the electroweak scale (standard model particles 
and Higgses).

We will be interested only in effective interactions that vanish or 
are suppressed at tree-level by a factor of $\tan\beta$ relative to 
their natural size, and we shall compute them at one loop only at leading
order in $1/\tan\beta$. In this case ${\cal G}_{\rm light}$ is 
suppressed relative to ${\cal G}_{\rm heavy}$ and can be set to 
zero. This can be understood from the fact that these interactions 
are protected by the PQ-symmetry mentioned above. This symmetry 
is broken strongly by the $\mu$ parameter that enters the 
high-scale contributions $ {\cal G}_{\rm heavy}$, but only 
weakly in the low-energy contributions by the small 
$b$ (or rather $m_{12}^2$, but in one-loop diagrams there is no distinction) 
parameter, $b\sim M_A^2/\tan\beta$. 
The low-energy contributions ${\cal G}_{\rm light}$ all involve an 
insertion of $b$, or an explicit coupling to the small vev $v_d \sim 
M_{\rm EW}/\tan\beta$, and are therefore negligible. 
On the other hand, the contribution 
${\cal G}_{\rm heavy}$ from the superpartner particle loops is already 
encoded in the effective couplings of the 2HDM (\ref{twohdm}). We 
therefore conclude that the leading one-loop contributions to the 
tree-suppressed Higgs couplings in the MSSM are simply obtained by 
computing the physical Higgs fields with the general 
tree-level 2HDM Lagrangian (\ref{twohdm}). Subsequently, the specific 
expressions for the couplings $z_{12}$, 
$m_{12}^2$, $\delta \lambda_{5,6,7}$, and $\delta \tilde y_{b,t,\tau}$ given 
in the previous section are inserted.

\subsection{\boldmath Vacuum expectation value $v_d$}

We first derive a relation between the shift (counterterm) of 
the vev $v_d$ and $\tan\beta$ and the MSSM parameter $b$, which feeds 
into all other Higgs couplings. The tadpole counterterm is 
determined by requiring that the one-loop $H_d$ tadpole 
(1PI one-point function) vanishes:
\begin{equation}
\Gamma_{H_d} = t_d+ \Delta t_d + [\Gamma_{H_d}]_{0,\rm heavy} + 
 [\Gamma_{H_d}]_{0,\rm light} \stackrel{!}{=}0 .
\label{vevcondition}
\end{equation}
The tadpole coupling $t_d$ is given by (\ref{parameterrel}) and 
set to zero by our renormalization convention. The tadpole 
counterterm in terms of the shifts of the vevs is also obtained from 
(\ref{parameterrel}) and given by 
\begin{eqnarray}
\Delta t_d &=& \Delta\hat m_d^2 v_d + \hat m_d^2\Delta v_d + 
\frac{1}{4} (\Delta g_1^2+\Delta g_2^{2}) v_d 
(v_d^2-v_u^2)-\Delta b v_u - b\Delta v_u 
\nonumber\\
&&
+\,\frac{3}{4} (g_1^2+g_2^{2}) v_d^2 \Delta v_d
-\frac{1}{4} (g_1^2+g_2^{2}) v_u^2 \Delta v_d
-\frac{1}{2} (g_1^2+g_2^{2}) v_d v_u \Delta v_u.
\label{Dtd}
\end{eqnarray}
The (unrenormalized) SUSY loop contribution to $\Gamma_{H_d}$ follows 
from the linear term in 
the 2HDM Lagrangian (\ref{twohdm}) after shifting the neutral 
Higgs fields by the vevs, which leads to 
\begin{equation}
[\Gamma_{H_d}]_{0,\rm heavy}= 
\delta m_{11}^2 v_d -\delta b v_u+\delta \lambda_1 
v_d^3  - 3 \delta \lambda_6 v_d^2 v_u+
(\delta \lambda_3+\delta\lambda_4+\delta \lambda_5)  v_d v_u^2 
- \delta \lambda_7 v_u^3,
\label{dtd}
\end{equation}
where $\delta m_{11}^2$, $\delta \lambda_{1-4}$ denote the 
SUSY loop contributions to the corresponding 2HDM parameters, which, 
however, will not be needed later on. The low-energy 
contributions can be neglected, as explained above, so the 
desired relation for $\Delta v_d$ follows from 
$\Delta t_d + [\Gamma_{H_d}]_{0,\rm heavy}\stackrel{!}{=}0$.

This condition simplifies in the large-$\tan\beta$ limit. 
Applying $v_d\ll v_u$, $b\ll M_{\rm EW}^2$, (\ref{Dtd}) reads 
\begin{equation}
\Delta t_d \approx \hat m_d^2\Delta v_d 
-\Delta b v_u -\frac{1}{4} (g_1^2+g_2^{2}) v_u^2 \Delta v_d 
\approx M_A^2\Delta v_d -\Delta b \,v
\label{Dtdsimp}
\end{equation}
The second approximation uses
\begin{eqnarray}
g_1^{2} &=& \frac{2 (M_Z^2-M_W^2)}{v^2},
\quad
g_2^2 = \frac{2 M_W^2}{v^2},
\nonumber\\
b &=& \frac{1}{2} M_A^2 \sin(2\beta), 
\nonumber\\
\hat m_d^2&=& \frac{1}{2} M_A^2 \sin(2\beta)\tan\beta-\frac{1}{2} 
M_Z^2 \cos(2\beta) \approx M_A^2+\frac{M_Z^2}{2},
\nonumber\\
\hat m_u^2&=& \frac{1}{2} M_A^2 \sin(2\beta)\cot\beta+\frac{1}{2} 
M_Z^2 \cos(2\beta) \approx -\frac{M_Z^2}{2},
\nonumber\\
v_d &=& v \cos\beta,
\quad
v_u = v \sin\beta \approx v,
\end{eqnarray}
which follow from inverting (\ref{parameterrel}) for the renormalized 
parameters together with $t_u=t_d=0$. Similarly, 
\begin{equation}
[\Gamma_{H_d}]_{0,\rm heavy} \approx 
-\delta b \,v - \delta \lambda_7 v^3.
\label{dtdsimp}
\end{equation}
Solving for $\Delta v_d$ results in 
\begin{equation}
\Delta v_d = \frac{(\delta b+\Delta b) v + \delta \lambda_7 v^3}{M_A^2}.
\label{vevresult}
\end{equation}
We see that unless the counterterm $\Delta b$ is fine-tuned the 
shift of the vev is of order $M_{\rm SUSY}^2/M_{\rm EW}$, which 
is very large relative to its assumed value of order 
$M_{\rm EW}/\tan\beta$. As discussed above, the shift of the vev is a finite 
quantity at leading order in large $\tan\beta$, since $\Delta b$ cancels the 
divergences in $\delta b$. Similar statements now 
apply to the shift of $\tan\beta$. Inserting (\ref{vevresult}) 
into (\ref{tanbcounter}), we find 
\begin{equation}
\Delta t_\beta \approx - \tan^2\beta\,\frac{\Delta v_d}{v} 
= - \tan^2\beta\,\frac{\delta b+\Delta b + \delta \lambda_7 v^2}
{M_A^2} +O(\tan\beta).
\label{tanbcounter2}
\end{equation}
The Higgs masses themselves receive no $\tan\beta$-enhanced one-loop 
corrections, so $M_A$ corresponds to the physical 
mass of the CP-odd Higgs boson even at the one-loop level in 
our approximations. 

\subsection{\boldmath Renormalization of $\tan\beta$ }

Contrary to the other parameters of the second set, there is no 
natural renormalization convention for the parameter 
$\tan\beta$. Gauge-independence and numerical stability of 
various schemes have been investigated 
in~\cite{Freitas:2002um,Baro:2008bg}. Quite generally, we 
expect problems with stability of observables 
against adding perturbative corrections, if the finite 
part of a counterterm is large, since this leads to a large shift 
between the tree-level and one-loop corrected values of 
parameters and large counterterm contributions 
to observables.

From (\ref{tanbcounter2}) we conclude that $\tan\beta$ receives 
large relative corrections, i.e. 
$\Delta t_\beta/\tan\beta \propto \tan\beta$, unless 
\begin{equation}
\delta b+\Delta b+\delta\lambda_7 v^2 =O(1/\tan\beta).
\label{tbren}
\end{equation}
Some of the stability features at large $\tan\beta$ discussed 
in~\cite{Freitas:2002um,Baro:2008bg}
can be understood on the basis of this equation. The $\overline{\rm DR}$ 
scheme for $\tan\beta$ has $\Delta t_\beta=0$ by definition and 
fulfills (\ref{tbren}) as does the DCPR 
definition~\cite{Dabelstein:1994hb,Chankowski:1992er}, for which 
$\Delta t_\beta/\tan\beta \propto 1$. Indeed, these two schemes 
were found to exhibit good stability. On the other hand, the 
$m_3$-scheme ($\overline{\rm DR}$ for the $b$ parameter), implies 
\begin{equation}
\frac{\Delta t_\beta}{\tan\beta} \approx 
- \tan \beta\,\frac{[\delta b]_{\rm fin}+\delta\lambda_7 v^2}
{M_A^2}, 
\end{equation}
which is very large, of order $M_{\rm SUSY}^2/M_{\rm EW}^2\times 
\tan\beta$. The Higgs mass and tadpole scheme~\cite{Freitas:2002um} 
also violate (\ref{tbren}), and are unstable. 

Good perturbative renormalization schemes should thus 
have $\delta b+\Delta b+\delta\lambda_7 v^2=0$ in the large-$\tan\beta$ 
limit. Since $b\sim M_{\rm EW}^2/\tan\beta$ but 
$\delta b \sim M_{\rm SUSY}^2$, this implies fine-tuning of the 
counterterm for the $b$ parameter as expected. 

\subsection{Physical Higgs fields}

\begin{figure}[!t]
  \begin{center}
  \includegraphics[width=0.3\textwidth]{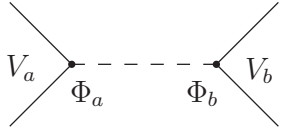}
  \caption{Virtual Higgs boson exchange.}
  \label{fig3}
  \end{center}
\end{figure}

To determine the physical Higgs fields, we first bring 
the kinetic terms of the 2HDM~(\ref{twohdm}) into
canonical form by a field redefinition. This is not 
necessary to determine the Higgs masses, which receive no 
$\tan\beta$-enhanced corrections, since there are no tree-level 
suppressions, or observables that involve only 
highly virtual internal Higgs lines, as for instance 
in low-energy flavour physics. The latter statement can be illustrated 
with the help of  Figure~\ref{fig3}, representing part of 
a diagram with an internal Higgs propagator. The expression for 
this diagram using the Lagrangian~(\ref{twohdm}) is 
$V_b^\dagger \,i \,[(Z q^2 -m^2)^{-1} ]_{ba} V_a$, where the 
propagator is a $2\times 2$ matrix in Higgs doublet space. 
In low-energy scattering the momentum transfer satisfies  
$q^2\ll m^2$, so the term involving the kinetic-mixing matrix $Z_{ab}$ 
in the 2HDM Lagrangian drops out at leading order in 
the low-energy expansion, and cannot affect physical observables. 
If one performs a field rotation to diagonalize $Z_{ab}$, the 
rotation affects the vertices $V_a$ and mass matrix $m_{ab}^2$ 
in such a way, that once again the effect disappears at 
leading order in $q^2/m^2$. However, in this paper 
our aim is to calculate the couplings of on-shell Higgs bosons 
($q^2\sim m^2$) to quarks and the Higgs fields themselves, relevant to 
Higgs production and decay, so we need the 
fields corresponding to the physical states. 

The field redefinition that diagonalizes the kinetic terms 
(and includes the shift of the fields by the vevs (\ref{introducevevs})) is 
\begin{eqnarray}
\Phi_1^1 = H_d^1 &\to& [v_d]_0+ \left(1-\frac{z_{11}}{2}\right) H_d^0+
\frac{1+a}{2} z_{12} H_u^{0*},
\nonumber\\
\Phi_1^2 = H_d^2 &\to& \left(1-\frac{z_{11}}{2}\right) H_d^--
\frac{1+a}{2} z_{12} H_u^{+*},
\nonumber\\
-\Phi_2^{1*} = H_u^2 &\to& [v_u]_0+ \left(1-\frac{z_{22}}{2}\right) H_u^0+
\frac{1-a}{2} z_{12} H_d^{0*},
\nonumber\\
\Phi_2^{2*} = H_u^1 &\to& \left(1-\frac{z_{22}}{2}\right) H_u^+-
\frac{1-a}{2} z_{12} H_d^{-*},
\label{kineticredef}
\end{eqnarray}
where $z_{ab}$ is the one-loop correction to $Z_{ab}$, and we used 
that the hermitian matrix $Z_{ab}$ is also symmetric due to 
CP-conservation.
The diagonal coefficients $z_{11}$, $z_{22}$ can be set to zero, 
since they are ordinary one-loop 
corrections to a non-vanishing tree term. The interesting 
terms are those that mix $H_d$ with the complex conjugate of $H_u$. 
The arbitrary quantity $a$ parameterizes a real field rotation 
in $(\Phi_1,\Phi_2)$ space, which preserves the diagonal form 
of the kinetic term. We could set $a=0$, but prefer to keep it to 
demonstrate explicitly the independence of physical 
quantities on $a$ below. Note that we do not rotate the 
fields and then shift them by the vevs, since the vevs (and $\tan\beta$) 
have been defined as parameters of the MSSM Lagrangian  before 
matching to the 2HDM.

After substituting (\ref{kineticredef}) into (\ref{twohdm}), we perform 
a unitary (in fact orthogonal, on account of CP-conservation) field 
rotation to diagonalize the Higgs mass matrix. The transformation to 
the physical Higgs fields $h^0,H^0,A^0,H^\pm$, 
including the pseudo-Goldstone fields $G^0,G^\pm$, is
\begin{eqnarray}
\left(\begin{array}{c} \mbox{Im}\,H_u^0\\ \mbox{Im}\,H_d^0 
\end{array}\right) &=& 
\frac{1}{\sqrt{2}} \left(\begin{array}{cc} 
\phantom{-}s_\beta+\delta s_\beta & c_\beta+\delta c_\beta\\
-[c_\beta+\delta c_\beta] & s_\beta+\delta s_\beta 
\end{array}\right) \,
\left(\begin{array}{c} G^0\\ A^0 
\end{array}\right),
\nonumber\\[0.2cm]
\left(\begin{array}{c} H_u^+\\ H_d^{-*} 
\end{array}\right) &=& 
\left(\begin{array}{cc} 
\phantom{-}s_\beta+\delta s_\beta & c_\beta+\delta c_\beta\\
-[c_\beta+\delta c_\beta] & s_\beta+\delta s_\beta 
\end{array}\right) \,
\left(\begin{array}{c} G^+\\  H^+ 
\end{array}\right),
\nonumber\\[0.2cm]
\left(\begin{array}{c} \mbox{Re}\,H_u^0\\ \mbox{Re}\,H_d^0 
\end{array}\right) &=& 
\frac{1}{\sqrt{2}} \left(\begin{array}{cc} 
\phantom{-}c_\alpha+\delta c_\alpha & s_\alpha+\delta s_\alpha\\
-[s_\alpha+\delta s_\alpha] & c_\alpha+\delta c_\alpha\\
\end{array}\right) \,
\left(\begin{array}{c} h^0\\ H^0 
\end{array}\right),
\end{eqnarray}
where $\delta s_\beta,\delta c_\beta, \delta s_\alpha,
\delta c_\alpha$ parameterize the correction to 
the corresponding MSSM tree-level rotation, and we use 
the conventional notation $s_\phi\equiv \sin\phi$, $c_\phi\equiv 
\cos\phi$. We already incorporated here that the 
correction $\delta c_\beta$ to the tree-level mixing 
matrix turns out to be the same for the CP-odd and the 
charged Higgs fields. The mixing angle $\alpha$ is given by 
\begin{equation}
\tan 2\alpha = \frac{M_A^2+M_Z^2}{M_A^2-M_Z^2}\,\tan 2\beta.
\label{alphatree}
\end{equation}

The correction terms $\delta s_\beta$, $\delta c_\beta$ 
are of the size of an ordinary loop correction, and hence 
relevant only if the corresponding tree contribution is 
suppressed. This is the case for the off-diagonal elements, 
since $c_\beta \propto 1/\tan\beta$. We therefore neglect the 
$\delta s_\beta$ terms relative to $s_\beta\approx 1$. 
For the off-diagonal correction we obtain 
\begin{equation}
\delta c_\beta  = -\frac{1+a}{2}\,z_{12} + \frac{\delta b+\Delta
  b+\delta\lambda_7 v^2}{M_A^2}.
\label{dcbe}
\end{equation} 
The second term vanishes in ``good'' renormalization schemes.

In determining the correction to $\alpha$, the cases 
$M_A>M_Z$ and $M_Z>M_A$ should be distinguished. In the following 
we discuss explicitly only the case $M_A>M_Z$. The other case 
follows roughly (that is, up to some signs) 
from interchanging $h^0$ and $H^0$. For large $\tan\beta$ we have 
$\tan 2\beta \approx -2/\tan\beta$, so (\ref{alphatree}) implies 
that either $c_\alpha$ or $s_\alpha$ is small, unless 
$M_A$ is very close to $M_Z$ to compensate the $\tan\beta$ 
suppression. For $M_A>M_Z$ 
the relevant case is $s_\alpha$ small and negative and $c_\alpha$ 
near 1.  The correction to $\tan 2\alpha$ 
is (valid for both cases $M_A>M_Z$, and $M_A<M_Z$)
\begin{eqnarray}
\delta t_{2\alpha} &=& a z_{12} + \frac{M_A^2+M_Z^2}{M_A^2-M_Z^2}\,
z_{12} 
\nonumber\\
&& -\,\frac{2}{M_A^2}\left( \frac{M_A^2+M_Z^2}{M_A^2-M_Z^2}\,
\left[\delta b+\Delta b+\delta\lambda_7 v^2\right] + 
\frac{2 M_A^2}{M_A^2-M_Z^2}\,\delta\lambda_7 v^2\right).
\label{dt2al}
\end{eqnarray}
For $M_A>M_Z$ we have $\delta s_\alpha = 
\frac{1}{2}\delta t_{2\alpha}$, $\delta c_{\alpha} \approx 0$ (relative 
to $c_\alpha\approx 1$). 
The field redefinitions that lead to the mass eigenstates therefore read
\begin{eqnarray}
H_u^0 &=&\frac{1}{\sqrt{2}}\left(c_\alpha h^0+[s_\alpha+\delta s_\alpha]
\,H^0 +i [c_\beta+\delta c_\beta] \,A^0+i s_\beta G^0\right),
\nonumber\\
H_d^0 &=&\frac{1}{\sqrt{2}}\left(- [s_\alpha+\delta s_\alpha] \,h^0+
c_\alpha H^0 +i s_\beta A^0-i [c_\beta+\delta c_\beta] \,G^0\right),
\nonumber\\
H_d^- &=& s_\beta H^- - [c_\beta+\delta c_\beta]\, G^-,
\nonumber\\[0.2cm]
H_u^+ &=& [c_\beta+\delta c_\beta]\, H^+ +s_\beta  G^+.
\label{physfields}
\end{eqnarray}

We briefly comment on the case $M_A=M_Z$, since (\ref{dt2al}) appears 
to be singular for this parameter input. The masses used here are 
tree-level masses, since we keep loop corrections only when they 
are enhanced by $\tan\beta$ relative to the tree term. When $M_A=M_Z$ 
the CP-even Higgs bosons $h^0$, $H^0$ are mass-degenerate and in 
fact physically indistinguishable. Eq.~(\ref{dt2al}) is derived under 
the assumption that the difference between the diagonal elements of 
the neutral Higgs boson mass matrix is not small. Otherwise one must 
keep the one-loop correction to the diagonal elements in the 
calculation of $\delta t_{2\alpha}$, and the approximation discussed 
here cannot be used. Thus the resonant enhancement 
in (\ref{dt2al}) is a real effect as long as 
$M_A^2-M_Z^2$ is large compared to $M_{\rm EW}^2 \times 
\mbox{ loop factor}$. However, if the widths of the Higgs bosons are so large 
that the $h^0$ and $H^0$ intermediate states are not separated by some 
measurement, nothing distinguishes the large tree-unsuppressed 
$c_\alpha H^0$ component 
from the smaller $[s_\alpha+\delta s_\alpha] h^0$ component 
in (\ref{physfields}) and there is no observable $\tan\beta$-enhancement 
in this measurement. In the following, we do not consider this 
degenerate case further.

\subsection{Yukawa couplings}

The Yukawa couplings and quark mass terms 
follow from inserting (\ref{physfields}) 
into (\ref{twohdm}). The bottom mass term is of special interest. 
The relevant bilinear terms are
\begin{equation}
{\cal L}_{b,\rm mass} = 
- \Big[(y_b+\delta y_b+\Delta y_b) (v_d+\Delta v_d) + 
\delta \tilde y_b (v_u+\Delta v_u)\Big] \,\bar b_R b_L + \mbox{h.c.}
\label{massb}
\end{equation}
The physical (pole) quark mass $m_b$ is obtained from this mass term 
plus low-energy self-energy contributions calculated in the 
2HDM. As before we keep one-loop corrections only when they 
are enhanced relative to the tree expressions. This allows us 
to neglect the low-energy contributions 
(see~\cite{Hempfling:1994ar,Kniehl:1991ze} for the standard model 
electroweak loops) 
as well as $\Delta v_u$, and $\delta y_b$ in (\ref{massb}). 
$\Delta y_b$ is the counterterm that cancels the divergences 
from superpartner loops in $\delta y_b$. The finite part of 
$\Delta y_b$ defines the relation between the physical bottom 
quark mass and the renormalized Yukawa coupling $y_b$. 
There is no reason to artificially define this finite part 
to be parametrically larger than the one of $\delta y_b$, so we may 
drop $\Delta y_b$ as well. Introducing the definitions
\begin{equation}
\epsilon_t=\frac{\delta \tilde y_t}{y_t},\quad
\epsilon_b=\frac{\delta \tilde y_b}{y_b},\quad
\epsilon_\tau=\frac{\delta \tilde y_\tau}{y_\tau}
\end{equation} 
we obtain the mass terms 
\begin{equation}
{\cal L}_{\rm mass} = 
-m_t \bar t_R t_L - m_b \bar b_R b_L - m_\tau \bar \tau_R \tau_L+ \mbox{h.c.}
\label{mass}
\end{equation}
with 
\begin{eqnarray}
m_t &=& y_t v_u = y_t v s_\beta,
\nonumber\\
m_b &=& y_b (v_d+\Delta v_d) + 
\delta \tilde y_b v_u = y_b v  s_\beta \left(\frac{1}
{\tan\beta+\Delta t_\beta}+\epsilon_b \right).
\label{mb}
\end{eqnarray}
Note that we do not expand in $\Delta t_\beta$, since according 
to (\ref{tanbcounter2}) it counts as a
$\tan\beta$-enhanced loop correction relative to 
$\tan\beta$, just as $\epsilon_b\tan\beta$ does relative to 1. No 
such enhancements are present for the top quark mass, which 
retains its tree expression. The expression for $m_\tau$ is 
obtained from the one for $m_b$ by replacing $(y_b,\epsilon_b)
\to (y_\tau,\epsilon_\tau)$.

Solving (\ref{mb}) for $y_b$ and eliminating $y_{b,t}$ by 
$m_{b,t}$, as it is conventionally done, we obtain the Higgs-Yukawa 
interactions 
\begin{eqnarray}
{\cal L}_{\rm Yuk} &=& 
- \frac{m_t}{\sqrt{2} v}\Bigg\{
\left(\frac{1}{\tan\beta} +\epsilon_t +\delta c_\beta 
-\frac{1-a}{2} z_{12}\right) 
\left(i\,\bar t_R t_L A^0-\sqrt{2} \bar t_R b_L H^+\right)
\nonumber\\
&&\hspace*{1.7cm} + \,\frac{c_\alpha}{s_\beta} \,\bar t_R t_L h^0
+ \frac{s_\alpha + \delta s_\alpha - 
\epsilon_t+\frac{1-a}{2} z_{12}}{s_\beta} \,\bar t_R t_L H^0
\nonumber\\
&&\hspace*{1.7cm} + \,i \,\bar t_R t_L G^0-\sqrt{2} \bar t_R b_L G^+
\Bigg\}
\nonumber\\
&& - \,\frac{m_b \,[\tan\beta+\Delta t_\beta]}
{\sqrt{2} v \,(1+\epsilon_b \,[\tan\beta+\Delta t_\beta])}
\Bigg\{i\,\bar b_R b_L A^0-\sqrt{2} \bar b_R t_L H^-
\nonumber\\
&&\hspace*{1.7cm}-\,\frac{s_\alpha + \delta s_\alpha - 
\epsilon_b-\frac{1+a}{2} z_{12}}{s_\beta} \,\bar b_R b_L h^0
+ \frac{c_\alpha}{s_\beta} \,\bar b_R b_L H^0
\nonumber\\
&&\hspace*{1.7cm} - \,
\left(\frac{1}{\tan\beta} +\epsilon_b +\delta c_\beta 
+\frac{1+a}{2} z_{12}\right) 
\left(i\,\bar b_R b_L G^0-\sqrt{2} \bar b_R t_L G^-\right)
\Bigg\}
\nonumber\\
&& + \, {\rm h.c.},
\label{eq:yukawa}
\end{eqnarray}
where as before we neglect ordinary 
one-loop corrections relative to unsuppressed tree terms, and 
the $\tau$ Yukawa couplings are given by the bottom ones after replacing 
all bottom parameters and fields by the corresponding expressions for 
tau. Several comments can be made. First, the explicit dependence on 
the arbitrary 
parameter $a$ cancels with the implicit dependence in 
(\ref{dcbe}), (\ref{dt2al}). Second, our result is independent on 
the renormalization convention for $\tan\beta$. A change of 
renormalization prescription is effected by a change in the finite 
part of $\Delta t_\beta$, or (\ref{tanbcounter2}) $\Delta v_d$, or 
(\ref{vevresult}) $\Delta b$. From (\ref{dcbe}), (\ref{dt2al}), we 
therefore find that 
\begin{equation}
[\tan\beta+\Delta t_\beta], 
\left[\frac{1}{\tan\beta} + \delta c_\beta\right], 
[s_\alpha + \delta s_\alpha], 
s_\beta,c_\alpha
\end{equation}
are invariant under a change of renormalization prescription 
that contains only $\tan\beta$-enhanced one-loop terms. 

Eq.~(\ref{eq:yukawa}) shows that a resummation of $\tan\beta$-enhanced 
terms is necessary only in the overall factor multiplying the bottom 
(and tau) Yukawa couplings~\cite{Carena:1999py}, where the need for 
resummation arises from using the quark (lepton) mass as a parameter 
rather than the MSSM Yukawa coupling. All of the Yukawa couplings 
which are $\tan\beta$-suppressed at tree-level receive unsuppressed 
loop corrections, which are incorporated in~(\ref{eq:yukawa}) 
through the ``wrong-Higgs'' Yukawa couplings $\epsilon_{t,b,\tau}$,  
and the quantities $\delta c_\beta$, 
$\delta s_\alpha$, $z_{12}$ from the one-loop effects in the 
Higgs sector. The loop-induced couplings may well exceed the tree terms, 
if $\tan\beta$ is large. 

To compare our result with previous results in the literature, 
see Eq.~(21) in \cite{Carena:1999py}, 
we consider the bottom quark 
coupling to the lightest Higgs boson $h^0$, which is the product 
of two factors
\begin{equation}
\frac{m_b \,[\tan\beta+\Delta t_\beta]}
{\sqrt{2} v \,(1+\epsilon_b \,[\tan\beta+\Delta t_\beta])}
\times \frac{s_\alpha + \delta s_\alpha - 
\epsilon_b-\frac{1+a}{2} z_{12}}{s_\beta}.
\end{equation}
With respect to the first factor, note that we do not discuss 
logarithmic one-loop effects, and hence do not distinguish the 
physical quark mass $m_b$ from the running $\overline{\rm MS}$ 
quark mass $\overline{m}_b(Q)$ used in \cite{Carena:1999py}. In reality, 
renormalization group evolution from $m_b$ to the electroweak scale 
is an important effect that should be taken into account as 
done in \cite{Carena:1999py}. The main difference in the first 
factor constitutes the presence of the (finite) counterterm 
$\Delta t_\beta$, which must not be neglected in general, and which 
renders the resummed result manifestly scheme-independent. This 
clarifies that the previously known result where $\Delta t_\beta=0$, 
is only valid in ``good'' renormalization schemes (\ref{tbren}) 
where the one-loop shift of $\tan\beta$ is not 
itself $\tan\beta$-enhanced, as also observed recently in the
related context of the muon Yukawa coupling \cite{Marchetti:2008hw}. 
The resummation formula 
from \cite{Carena:1999py} without the counterterm contribution
should not be used in other schemes, for which  
$\Delta t_\beta$ is an important correction in the first factor.

Using $s_\beta \approx 1$ in the one-loop corrections, and 
(\ref{tanbcounter2}), (\ref{dt2al}), the second factor can be 
rewritten as 
\begin{equation}
\frac{s_\alpha}{s_\beta} - \epsilon_b + 
\frac{M_Z^2}{M_A^2-M_Z^2}\, z_{12} - 
\frac{2 \delta\lambda_7 v^2}{M_A^2-M_Z^2} 
-\frac{M_A^2+M_Z^2}{M_A^2-M_Z^2}\,
\frac{\delta b+\Delta b+\delta\lambda_7 v^2}{M_A^2}.
\label{eq:bbh0}
\end{equation}
The first two terms reproduce the corresponding result 
in \cite{Carena:1999py}, but the Higgs-sector effects 
embodied in the last three terms have not been included 
there. Higgs-sector effects related to kinetic mixing have 
recently been computed in \cite{Freitas:2007dp}. The 
$\overline{\rm DR}$ scheme for $\tan\beta$ is assumed 
there, hence the last term involving $\delta b+\Delta b+\delta\lambda_7 v^2 
\propto \Delta t_\beta=0$ vanishes. We find several discrepancies
in the comparison with~\cite{Freitas:2007dp}.
The third term proportional to $z_{12}$ would be consistent with the result of 
\cite{Freitas:2007dp} if $z_{12}=\epsilon_{\rm GP}$, but our 
result for $z_{12}$ in (\ref{z121loop}) gives 
$z_{12}=-\epsilon_{\rm GP}$. We also find that the corrections 
to the  $\bar{t}_L t_R H^0$ coupling are not included 
in~\cite{Freitas:2007dp}. The most important difference is that the 
fourth term involving $\delta\lambda_7$ is missed, since it is assumed 
that Higgs-sector effects arise only from kinetic mixing, and not (also) from 
the Higgs self-couplings. The large quantum correction 
to the tree-level suppressed charged Higgs $\bar t_R b_L H^+$ coupling 
is present in the discussion of the $b\to s\gamma$ transition in
\cite{Carena:2000uj,Ciuchini:1998xy}. These papers account for the 
gluino contribution to $\epsilon_t$. The Higgs-sector terms 
involving $\delta c_\beta$ and $z_{12}$ are again not considered.

More precisely, in the expression for the third term 
proportional to $z_{12}$ Ref.~\cite{Freitas:2007dp} 
has $M_{h^0}^2/(M_{H^0}^2-M_{h^0}^2)$, where we have 
$M_Z^2/(M_A^2-M_Z^2)$. The two expressions are consistent with each 
other, since in the one-loop correction we may use the 
large-$\tan\beta$ limits of the tree-level mass relations, which 
give $M_{h^0}=M_Z$, and $M_{H^0}=M_A$. In reality, the Higgs masses 
must receive large quantum corrections to satisfy experimental limits, 
which are not included in 
our equations, since they are not $\tan\beta$-enhanced. Since 
the apparent singularity for $M_A=M_Z$ arises from the 
degeneracy of the two CP-even Higgs bosons, as discussed above, 
it is plausible that 
$M_{h^0}^2/(M_{H^0}^2-M_{h^0}^2)$ provides a better approximation, 
even though formally the difference to $M_Z^2/(M_A^2-M_Z^2)$ 
is a higher-order effect. 

In Section~\ref{numestimate} we 
perform some representative numerical estimates of the new 
Higgs-sector corrections to the Yukawa couplings and compare them 
to the previously known term $\epsilon_b$.

\subsection{Higgs self-couplings}

The Higgs self-couplings are obtained in the same straightforward 
fashion. Here we focus on the trilinear couplings, since the quartic 
self-interactions with coefficients induced by quantum effects will 
be too small to be measured in any foreseeable experiment. The 
detection of $\tan\beta$- or loop-suppressed trilinear couplings 
provides a challenge even to the ILC (see, e.g., \cite{Djouadi:1999gv}).

The Lagrangian for the trilinear couplings 
\begin{equation}
{\cal L}_{HHH} = {\cal L}_{HHH,\,\rm large} + 
{\cal L}_{HHH,\,\rm small}
\end{equation}
consists of a piece ${\cal L}_{HHH,\,\rm large}$ involving 
interactions that do not vanish at tree level when 
$\tan\beta\to\infty$, and another piece ${\cal L}_{HHH,\,\rm small}$ 
with interactions that do. The tree-level couplings in the MSSM are 
standard and will not be repeated here. We find a compact expression 
for the $\tan\beta$-enhanced 
one-loop corrections relative to the small tree-level interactions in 
${\cal L}_{HHH,\,\rm small}$, applicable to the case $M_A>M_Z$:
\begin{eqnarray}
{\cal L}_{HHH,\,\rm small}^{\rm one-loop} &=&
-\frac{v}{\sqrt{2}} \,\bigg\{
(-1)(3 M_A^2+2 M_Z^2) \,C_H \,{h^{0}}^2 H^{0} \nonumber\\[0.0cm]
&&  \hspace*{-1cm}
+ \,(M_A^2 \,C_H -\delta\lambda_6-\delta\lambda_7) \,({H^{0}}^2+{A^{0}}^2)H^{0}
\nonumber\\[0.1cm]
&&  \hspace*{-1cm}
+ \,2\,\Big[(M_A^2-2 M_W^2) \,C_H -(\delta\lambda_6+\delta\lambda_7)\Big] 
\,H^0 H^+ H^- 
\nonumber\\[0.1cm]
&&  \hspace*{-1cm}
+ \, 2 \,(M_A^2-M_Z^2)\,C_H \,h^0 A^0 G^0 - 2 \delta\lambda_5 \,H^0 A^0 G^0 
- M_A^2\,C_H\,H^0 ({G^0}^2 + 2\,G^+ G^-)
\nonumber\\[0.1cm]
&& \hspace*{-1cm} 
+ \,2\,(M_A^2-M_Z^2+M_W^2) \,C_H\,h^0 (H^- G^++H^+ G^-)
\bigg\}\,,
\label{eq:tri}
\end{eqnarray}
where
\begin{equation}
C_H = \frac{1}{M_A^2-M_Z^2} \left(
\frac{M_Z^2}{v^2}\left[\frac{\delta b+\Delta b+\delta\lambda_7 v^2}{M_A^2} - 
\frac{z_{12}}{2}\right] + \delta\lambda_7 \right).
\end{equation}

\section{Numerical estimates}
\label{numestimate}

In this section we present numerical estimates of the Higgs-sector 
corrections by evaluating them for degenerate SUSY masses, and by 
performing a random parameter space scan. We are especially interested 
in comparing the new Higgs-sector effects to the known one-loop 
induced ``wrong-Higgs'' Yukawa couplings.

\subsection{Degenerate SUSY masses}

We begin the analysis of the size of the new terms in the Yukawa couplings 
relative to $\epsilon_{b,t}$ by assuming that all 
SUSY parameters are equal to a 
common mass scale $M_{\rm SUSY}$ and by allowing for arbitrary signs in 
the $\mu$ and $A$ parameters, but not the gaugino masses. In the following 
we give analytic results for  $\epsilon_{b,t,\tau}$ and the quantities 
$z_{12}$ and $\delta\lambda_{5-7}$, which determine the 
Higgs-sector corrections to the Yukawa couplings and the Higgs 
self-couplings, and numerical results to judge the importance 
of individual terms. To make contact with the literature, we 
introduce the rescaled $A$ parameters $\widetilde{A}$ by 
$A_i = y_i \widetilde{A}_i$. In units of ${\rm sgn}(\mu)/(96\pi^2)$, 
we find
\begin{eqnarray}
\epsilon_b &=& \, 3 y_t^2 \, {\rm sgn} (\widetilde{A}_t) + 32\pi \alpha_s 
- 18\pi \alpha_{\rm em} \left(\frac{1}{s^2_w} + \frac{11}{27 c^2_w} \right) 
\nonumber\\
&=& 9.65 \,+\, {\rm sgn}(\widetilde{A}_t) \, \frac{3.02}{s_\beta^2} \,,
\nonumber\\
\epsilon_t &=& -\, 3 y_b^2 \, {\rm sgn} (\widetilde{A}_b) - 32\pi \alpha_s 
+ 18\pi \alpha_{\rm em} \left(\frac{1}{s^2_w}+ \frac{5}{27 c^2_w} \right) 
\nonumber\\
&=& -9.77 \,-\, {\rm sgn} (\widetilde{A}_b) \, \frac{2.14}{[50\,c_\beta]^2} \,,
\nonumber\\
\epsilon_\tau &=& \,- 18\pi \alpha_{\rm em} 
\left(\frac{1}{s^2_w} - \frac{1}{3 c^2_w} \right) = -1.80
\label{eq:analytic1}
\end{eqnarray}
for the ``wrong-Higgs'' Yukawa couplings, and 
\begin{eqnarray}
z_{12} &=& 3 y_t^2 \, {\rm sgn} (\widetilde{A}_t) + 3 y_b^2 \, {\rm sgn} (\widetilde{A}_b) 
+ y_\tau^2 \, {\rm sgn} (\widetilde{A}_\tau) 
+ 4\pi \alpha_{\rm em} \left(\frac{3}{s^2_w} + \frac{1}{c^2_w} \right) 
\nonumber\\
&=& 1.45 \,+\, {\rm sgn}(\widetilde{A}_t) \, \frac{3.02}{s_\beta^2} 
\,+\, {\rm sgn}(\widetilde{A}_b) \, \frac{2.14}{[50\,c_\beta]^2} 
\,+\, {\rm sgn}(\widetilde{A}_\tau) \, \frac{0.26}{[50\,c_\beta]^2} 
\,,
\nonumber\\[0.2cm]
\delta\lambda_5 &=& {\rm sgn}(\mu) \,\bigg\{ -  3 y_t^4 - 3 y_b^4 - y_\tau^4  
+ 16\pi^2 \alpha_{\rm em}^2 \left( \frac{3}{s^4_w} + \frac{1}{c^4_w} +\frac{2}{c^2_w s^2_w} \right) \bigg\}
\nonumber\\
&=&  {\rm sgn}(\mu) \,\bigg\{ 0.712 \,-\, \frac{3.04}{s_\beta^4} 
\,-\, \frac{1.59}{[50\,c_\beta]^4}  \bigg\}
\,,
\nonumber\\[0.2cm]
\delta\lambda_6 &=&
15 \, y_b^4\, {\rm sgn} (\widetilde{A}_b) - 3 y_t^4 \, {\rm sgn} (\widetilde{A}_t) 
+ 5 y_\tau^4 \, {\rm sgn} (A_\tau)  
\nonumber\\[0.1cm]
&& - \, \frac{3\pi \alpha_{\rm em}}{c^2_w \, s^2_w} 
\left( -\,3 y_t^2 \, {\rm sgn} (\widetilde{A}_t) + 3 y_b^2 \, {\rm sgn} (\widetilde{A}_b) + y_\tau^2 \, {\rm sgn} (\widetilde{A}_\tau)  \right)
\nonumber\\
&& + \, 32 \pi^2 \alpha_{\rm em}^2 
\left( \frac{3}{s^4_w} + \frac{1}{c^4_w} + \frac{2}{s^2_w c^2_w} \right) 
\nonumber\\
&=& 1.42 \,-\, {\rm sgn}(\widetilde{A}_t) \left(\frac{3.04}{s^4_\beta}
\,-\,\frac{1.29}{s_\beta^2}\right)
\,-\, {\rm sgn}(\widetilde{A}_b) \left(\frac{0.91}{[50\,c_\beta]^2}
\,-\, \frac{7.62}{[50\,c_\beta]^4}\right)
\nonumber\\
&& 
\,-\, {\rm sgn}(\widetilde{A}_\tau) \left(\frac{0.11}{[50\,c_\beta]^2}
\,-\, \frac{0.34}{[50\,c_\beta]^4}\right)
\,,
\nonumber\\[0.2cm]
\delta\lambda_7 &=& 15 \, y_t^4 \, {\rm sgn} (\widetilde{A}_t) - 3 y_b^4 \, {\rm sgn} (\widetilde{A}_b) 
- y_\tau^4 \, {\rm sgn} (\widetilde{A}_\tau) 
\nonumber\\[0.1cm]
&& + \, \frac{3\pi \alpha_{\rm em}}{c^2_w \, s^2_w} 
\left( -3 y_t^2 \, {\rm sgn} (\widetilde{A}_t) + 3 y_b^2 \, {\rm sgn} (\widetilde{A}_b) + y_\tau^2 \, {\rm sgn} (\widetilde{A}_\tau)  \right)
\nonumber\\
&& + \, 32 \pi^2 \alpha_{\rm em}^2 
\left( \frac{3}{s^4_w} + \frac{1}{c^4_w} + \frac{2}{s^2_w c^2_w} \right) 
\nonumber\\
&=& 1.42 \,+\, {\rm sgn}(\widetilde{A}_t) \left(\frac{15.2}{s^4_\beta}
\,-\, \frac{1.29}{s_\beta^2}\right)
\,+\, {\rm sgn}(\widetilde{A}_b) \left( \frac{0.91}{[50\,c_\beta]^2}
\,-\, \frac{1.52}{[50\,c_\beta]^4}\right) \nonumber\\
&&
\,+\, {\rm sgn}(\widetilde{A}_\tau) \left( \frac{0.11}{[50\,c_\beta]^2}
\,-\, \frac{0.07}{[50\,c_\beta]^4}\right)
\label{eq:analytic}
\end{eqnarray}
for the Higgs-sector couplings. Our results for $\delta\lambda_{5-7}$ agree  
with the special case given in~\cite{Haber:1993an}, which assumed all 
squark masses equal to $M_{\rm SUSY}$ and neglected the gaugino and 
slepton contributions.  The numerical values in (\ref{eq:analytic}) 
have been evaluated with SM parameter input as follows: 
for the gauge boson masses we choose $M_W=80.398\,$GeV and 
$M_Z=91.188\,$GeV, the gauge couplings are fixed to 
$\alpha_{\rm em}=\alpha_{\rm em}^{\rm \overline{MS}}(M_Z)=1/127.9$ 
and $\alpha_s=\alpha_s^{\rm \overline{MS}}(M_Z)=g_3^2/(4\pi)=0.118$. The 
cosine of the weak mixing angle is given by $c_w=M_W/M_Z$.  For the 
top Yukawa coupling we use
$y_t=\overline{m}_t/(v s_\beta)\simeq 1.003/s_\beta$, 
where $\overline{m}_t=171.7\,$GeV and
$v=s_w M_W/\sqrt{2\pi\alpha_{\rm em}}\simeq 171.2\,$GeV. For the 
down-type and lepton Yukawa couplings we here use the tree-level relations
$y_b=\overline{m}_b/(v c_\beta)\simeq 0.0169/c_\beta$ and 
$y_\tau=\overline{m}_\tau/(v c_\beta)\simeq 0.0102/c_\beta$, 
where $\overline{m}_b=2.89\,$GeV and $\overline{m}_\tau=1746.24\,$MeV. 
All fermion masses correspond to the renormalized masses in the 
$\rm \overline{MS}$ scheme at $\mu=M_Z$~\cite{Xing:2007fb} rather 
than the pole masses.

\begin{figure}[t]
  \begin{center}
  \vspace*{0.cm}
  \includegraphics[width=1\textwidth]{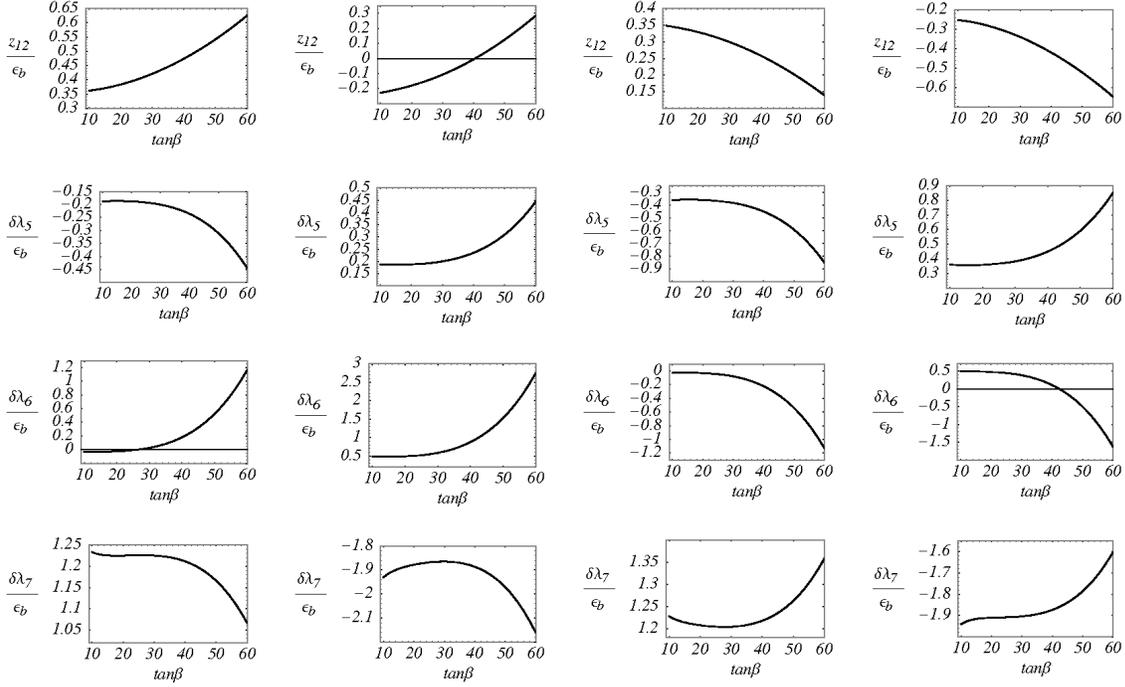}
  \vspace*{-0.4cm}
\caption{Dependence of Higgs-sector couplings, normalized to $\epsilon_b$, 
on $\tan\beta$. For the ratios 
$z_{12}/\epsilon_b$, $\delta\lambda_{6,7}/\epsilon_b$ 
the plots correspond to the sign choices 
$\{ {\rm sgn} (\widetilde{A}_t),{\rm sgn} (\widetilde{A}_b) \} = 
\{+,+\},\,\{-,+\},\,\{+,-\},\,\{-,-\}$. In the plots for 
$\delta\lambda_{5}/\epsilon_b$
we adopt $\{ {\rm sgn} (\mu),{\rm sgn} (\widetilde{A}_t) \} = 
\{+,+\},\,\{-,+\},\,\{+,-\},\,\{-,-\}$.}
  \label{fig4}
  \end{center}
\end{figure}

Since $\epsilon_b$ is nearly independent of $\tan\beta$ for 
$\tan\beta\gg 1$, we show the $\tan \beta$ dependence of the 
ratios $z_{12}/\epsilon_{b}$ 
and $\delta\lambda_{5-7}/\epsilon_{b}$ from (\ref{eq:analytic1}), 
(\ref{eq:analytic}) in Figure~\ref{fig4} for different choices 
of the signs of the $\widetilde{A}$-parameters. 
The slepton terms give a negligible contribution except for $z_{12}$ at 
very large values of $\tan\beta$, so we always set the sign of 
$\widetilde{A}_\tau$ to $+1$. For $\tan\beta$ smaller than 
20 -- 30  the value of $z_{12}$, $\delta\lambda_{5-7}$ is determined 
by the $\widetilde{A}_t$ terms and the gaugino contributions. 
For $\delta\lambda_7$, in  
particular, which feeds into the Yukawa couplings, the terms proportional 
to $\widetilde{A}_t$ give by far the largest contributions, while the 
effect of $\widetilde{A}_b$ becomes relevant for very large
$\tan\beta$ values, reaching up to a 20\% contribution.
On the other hand we observe from the plots that the $\widetilde{A}_b$ 
terms and a change in the sign of $\widetilde{A}_b$ has a large effect on 
$\delta\lambda_6$, which is driven by the 
${\rm sgn}(\widetilde{A}_b)/c_\beta^4$ term in (\ref{eq:analytic}) with its 
sizeable coefficient. In general, 
under the assumption of a common mass scale for all SUSY parameters, 
we thus conclude that the quantities $z_{12},\,
\delta\lambda_{5,6}$ and particularly 
$\delta\lambda_7$ are comparable in size with 
the ``wrong-Higgs'' couplings $\epsilon_{b,t}$ even though 
they do not receive gluino contributions. Note that $|\delta\lambda_7|$ is 
larger than $|\epsilon_{b,t}|$ for all choices of signs of the 
$\widetilde{A}_{t,b}$.

\subsection{Estimates using random 
parameter-space sampling}

We now investigate in more detail the size of the Higgs one-loop 
corrections to the Yukawa interactions relative to $\epsilon_{b,t}$ for 
different SUSY parameter input. We consider good renormalization 
schemes for $\tan\beta$, {\it i.e.} schemes satisfying 
$\delta b+\Delta b+\delta \lambda_7 v^2=0$ in the large-$\tan\beta$ 
limit, in which case the two relevant corrections from (\ref{eq:yukawa}) are
\begin{equation}
h_1 \equiv z_{12}, \qquad 
h_2 \equiv \frac{M_Z^2 z_{12} - 2v^2\delta\lambda_7}{M_A^2-M_Z^2} \,.
\label{eq:h1h2}
\end{equation}
In terms of the quantities $h_{1,2}$, the one-loop improved couplings 
of the Higgs fields to the bottom and top quark are rewritten as, omitting 
the global factors outside the curly brackets in Eq.~(\ref{eq:yukawa}),
\begin{eqnarray}
 \bar{b}_R b_L h^0 \quad &:& \quad \frac{s_\alpha}{s_\beta} +  
 \delta s_\alpha - \epsilon_b - \frac{1+a}{2} z_{12}
 = \frac{s_\alpha}{s_\beta} - \epsilon_b + h_2 \,,
 \nonumber\\
i \bar{b}_R b_L G^0 \quad &:& \quad 
\frac{1}{\tan\beta} + \epsilon_b + \delta c_\beta + \frac{1+a}{2} z_{12}
=  \frac{1}{\tan\beta} + \epsilon_b \,,
\nonumber\\
i\bar{t}_R t_L A^0 \quad &:& \quad \frac{1}{\tan\beta} + \epsilon_t + 
\delta c_\beta - \frac{1-a}{2} z_{12}
=  \frac{1}{\tan\beta} + \epsilon_t - h_1 \,,
\nonumber\\
\bar{t}_R t_L H^0 \quad &:& \quad \frac{s_\alpha}{s_\beta} + 
\delta s_\alpha - \epsilon_t + \frac{1-a}{2} z_{12}
= \frac{s_\alpha}{s_\beta} - \epsilon_t + h_1 + h_2 \,.
\label{yukawah}
\end{eqnarray}
The function $C_H$, which appears in the one-loop corrections to the small
tree-level trilinear Higgs couplings (\ref{eq:tri}), also has a simple 
expression in terms of $h_{2}$, given by
\begin{equation}
C_H = -\frac{h_2}{2v^2}\,.
\end{equation}

We perform a random parameter sampling for the SUSY parameters 
in the following ranges: 500 GeV$\,\le \tilde{m}_i \le 5\,$TeV, with 
$\tilde{m}_i$ being the squark, slepton or gaugino masses, and 
500 GeV$\,\le |\mu|, | \widetilde{A}_t|, | \widetilde{A}_b|, 
| \widetilde{A}_\tau| \le 5\,$TeV. The different sign assignments 
for the parameters $\mu$, $\widetilde{A}_t$ and 
$\widetilde{A}_b$ are explored separately, but we fix 
${\rm sgn}(\widetilde{A}_\tau) = +1$. Since the $M_A$ dependence of 
$h_{2}$ comes in only through the prefactor $1/(M_A^2-M_Z^2)$,
we shall fix it to the reference value $M_A=200\,$GeV. Here contrary to 
the analytic expressions in the previous subsection, we correctly include 
the $\tan\beta$-enhanced loop correction to the down-type quark 
and lepton Yukawa couplings, {\it i.e.} $y_{b,\tau}=\overline{m}_{b,\tau}
/(v c_\beta)\times 1/(1+\epsilon_{b,\tau} \tan\beta)$ is used. We remark  
that not all points allowed in the parameter scan are physical, since 
vacuum stability requirements do not allow large $A$ terms relative 
to the scale of the SUSY particle masses.

\begin{figure}[t]
  \begin{center}
  \includegraphics[width=\textwidth]{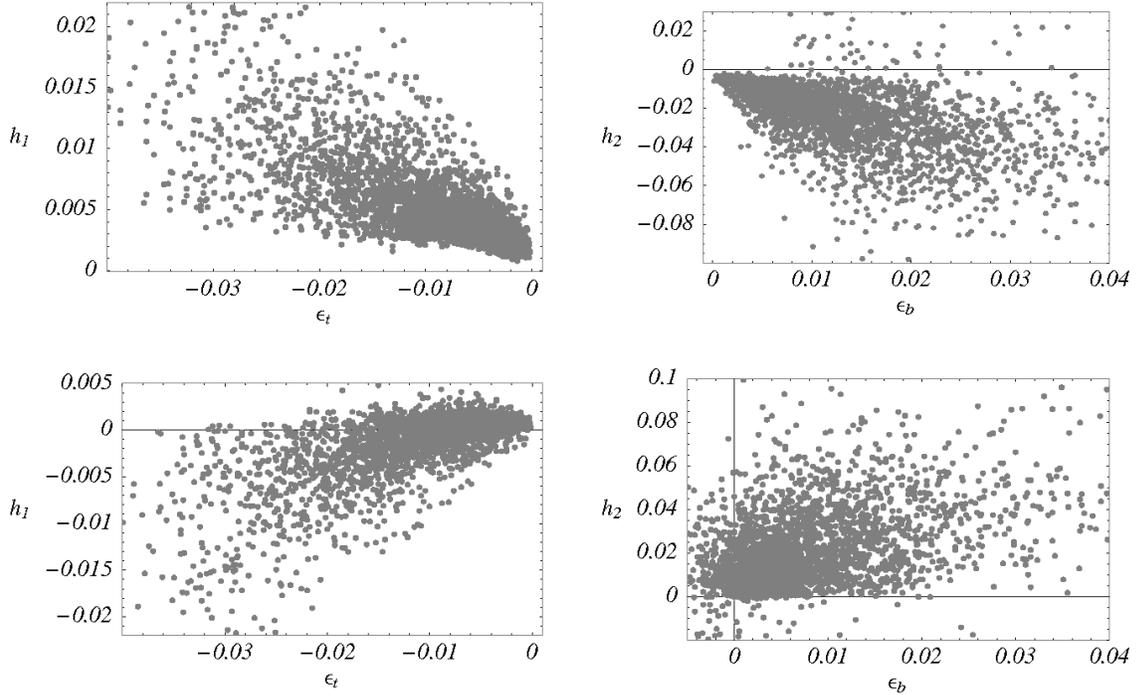}
  \caption{The corrections $h_1$ and $h_2$ versus $\epsilon_t$ and 
  $\epsilon_b$ respectively, for $M_A=200\,$GeV and positive $\mu$. 
  The plots in the first row have ${\rm sgn}(\widetilde{A}_t)=+1$ and 
  those in the second row have ${\rm sgn}(\widetilde{A}_t) = -1$
  (${\rm sgn}(\widetilde{A}_b)=+1$ in all plots).}
  \label{fig5}
  \end{center}
\end{figure}

Let us first study the case of positive $\mu$, where $\epsilon_b$ is 
preferentially positive, so that $y_b$ is reduced by the one-loop 
correction. In Figure~\ref{fig5} we show $h_1$ (left) and $h_2$ (right) 
versus $\epsilon_t$ and $\epsilon_b$, respectively, for both signs of 
$\tilde{A}_t$ (positive in the upper plots, negative in the lower ones). 
For positive $\mu$, the value of $\tan\beta$ does not significantly 
change the shape of the scatter plot, and hence has been fixed to 
$\tan\beta=35$. The typical size of the effective couplings is a few 
percent. Values of $|h_2|$ range up to 0.1, which implies that the 
one-loop induced coupling may exceed the $\tan\beta$-suppressed tree 
coupling for values of $\tan\beta$ as small as 10. In comparison, $h_1$, 
which originates only from kinetic mixing, tends to be significantly 
smaller. Thus, the Higgs sector correction 
$h_2$ competes in size with the ``wrong-Higgs'' Yukawa coupling 
$\epsilon_b$ even for 
large values of $M_A$, and is in fact the dominant one-loop correction 
to the $\bar{b}b h^0$ Yukawa interaction.
Some general features that can be observed in Figure~\ref{fig5}, 
or extracted from the analytic expressions are:

\begin{itemize}

\item[i/] In the assumed range of SUSY parameters, 
the sign of $\epsilon_b$ ($\epsilon_t$) is correlated (anti-correlated) 
with the sign of $\mu$, since the gluino contribution 
(first term in (\ref{eq:Deltayb})) dominates and the function $J_3$ is 
always negative. The term proportional to $\widetilde{A}_t$ can overcome 
the gluino contribution and change the sign of $\epsilon_b$ only 
for negative $\widetilde{A}_t$ in the parameter-space region  
where $|\widetilde{A}_t|/\mu^2$ is much larger than $1/M_3$. 
Similarly, large negative $\widetilde{A}_b$ can make $\epsilon_t$ 
positive, when $y_b^2|\widetilde{A}_b|/\mu^2 
\gg 1/M_3$.

\item[ii/] Since $\epsilon_b$ cannot reach large negative values, 
there is no significant enhancement of the bottom Yukawa coupling 
from the large-$\tan\beta$ resummation. The numerical effect of 
the terms proportional to $y^2_b \widetilde{A}_b$ is therefore small 
compared to those proportional to $y_t^2 \widetilde{A}_t$. We verified 
that the scatter plots for negative
$\widetilde{A}_b$ are similar to the ones with positive 
$\widetilde{A}_b$ shown in Figure~\ref{fig5}.

\item[iii/] Larger values of $|\epsilon_t|$ ($|\epsilon_b|$) tend to 
be correlated with larger values of $|h_1|$ ($|h_2|$). However, there 
is no strict relation, since the dominant contribution to the 
``wrong-Higgs'' Yukawa couplings is proportional to the gluino mass, 
while those to the Higgs-sector couplings involve the 
$A$ terms.

\item[iv/] The largest contribution to $h_1$ is given by the term 
proportional to $\widetilde{A}_t$ in~(\ref{z121loop}). However, the sign 
of $h_1$ is not determined uniquely by the sign of $\widetilde{A}_t$ due to 
the gaugino term proportional to $g_2^2$, which is always positive 
(the function $H_2$ is positive) and can become comparable to the 
$\widetilde{A}_t$ term in some regions of the parameter space.

\item[v/] In most of the explored parameter space the sign of $h_2$ is 
opposite to the sign of $\widetilde{A}_t$. Since the 
$\delta\lambda_7$ term 
in the definition (\ref{eq:h1h2}) of $h_2$ is the larger of the two, 
we have ${\rm sgn}(h_2)=-{\rm sgn}(\delta\lambda_7)$. The relation 
${\rm sgn}(\delta \lambda_7)={\rm sgn}(\widetilde{A}_t)$ can be understood as 
follows: neglecting the smaller terms involving gauge couplings, 
Eq.~(\ref{eq:lambda7}) for $\delta\lambda_7$ contains the terms
\begin{eqnarray}
&&-\, 3\mu\, y_t^4 \widetilde{A}_t \left\{
J_2(m_{\tilde Q},m_{\tilde t_R}) +
J_2(m_{\tilde t_R},m_{\tilde Q}) \right\}
-\, 3 \mu \,\widetilde{A}_t^3 y_t^4 K_2(m_{\tilde Q},m_{\tilde t_R})
\label{eq:At}
\end{eqnarray}
proportional to powers of $\widetilde{A}_t$. The first term has the same 
sign as $\widetilde{A}_t$, since  $J_2$
is always negative, while the second term has opposite sign, since 
$K_2$ is positive. Defining the ratio 
\begin{equation}
r(x) =  -\frac{J_2(m_{\tilde Q},m_{\tilde t_R}) + 
J_2(m_{\tilde t_R},m_{\tilde Q})}
{m^2_{\tilde Q}\,K_2(m_{\tilde Q},m_{\tilde t_R})}
\end{equation}
with  $x=m_{\tilde t_R}^2/m_{\tilde Q}^2$, we find that 
$r(x)$ is monotonically
increasing, satisfying $1.6 < r(x) < 169$ when $1/100<x<100$ as 
allowed by our parameter-space sampling. The third term in (\ref{eq:At}) 
can compete with the first only if 
$(\widetilde{A}_t/m_{\tilde{Q}})^2 > r(x)$. For $x=1$ this relation
requires large $A$ terms, $(\widetilde{A}_t/m_{\tilde{Q}})^2 > 6$, 
which are disfavoured 
by vacuum stability arguments. We also note that the largest gaugino 
term in $\delta\lambda_7$ (the one proportional to $g_2^4$)
gives a positive contribution because the function $L_2$ is negative. 

\end{itemize}

\begin{figure}[t]
  \begin{center}
  \includegraphics[width=\textwidth]{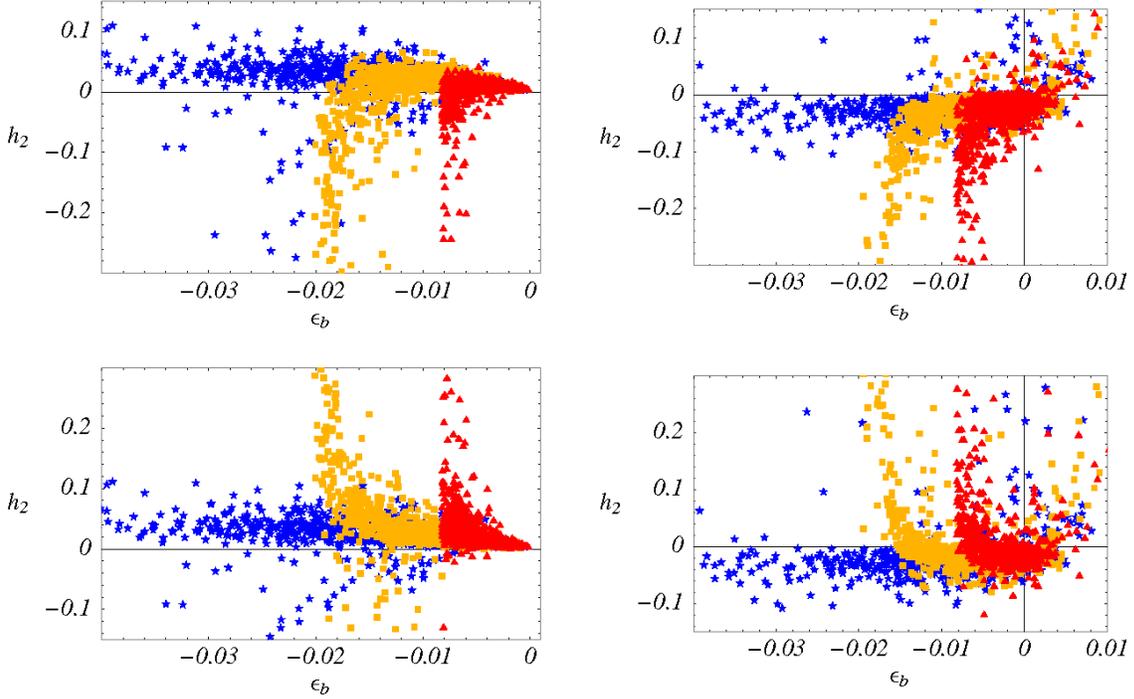}
  \caption{The correction $h_2$ versus $\epsilon_b$, for $M_A=200\,$GeV 
  and negative $\mu$. The plots in the first row correspond to 
  $\{ {\rm sgn} (\widetilde{A}_t), {\rm sgn} (\widetilde{A}_b) \}= 
  \{+1,+1\}$ and $\{-1,+1\}$,
  while those in the second row have 
  $\{ {\rm sgn} (\widetilde{A}_t), {\rm sgn} (\widetilde{A}_b) \}= 
  \{+1,-1\},\,\{-1,-1\}$.
  Black, lightgray and darkgray points (blue, yellow and red in colored plots) 
  correspond to
  $\tan\beta=10,\,35$ and $60$, respectively.}
  \label{fig6}
  \end{center}
\end{figure}

Let us now turn to the case ${\rm sgn}(\mu)=-1$. The most important 
difference is that the negative sign of the $\mu$ parameter flips the 
sign of $\epsilon_b$ to negative values (and the one of $\epsilon_t$ to 
positive values), which leads to a strong increase of the bottom 
and $\tau$ Yukawa coupling, when $\epsilon_b$ cancels the 
$1/\tan\beta$ term in (\ref{mb}). The validity of a perturbative 
expansion is in doubt when the Yukawa couplings become too large. 
In the parameter space sampling for negative $\mu$, we therefore 
only keep points that satisfy $0<y_{b,\tau}<2$.

\begin{figure}[t]
  \begin{center}
  \vspace*{-0.1cm}
  \includegraphics[width=\textwidth]{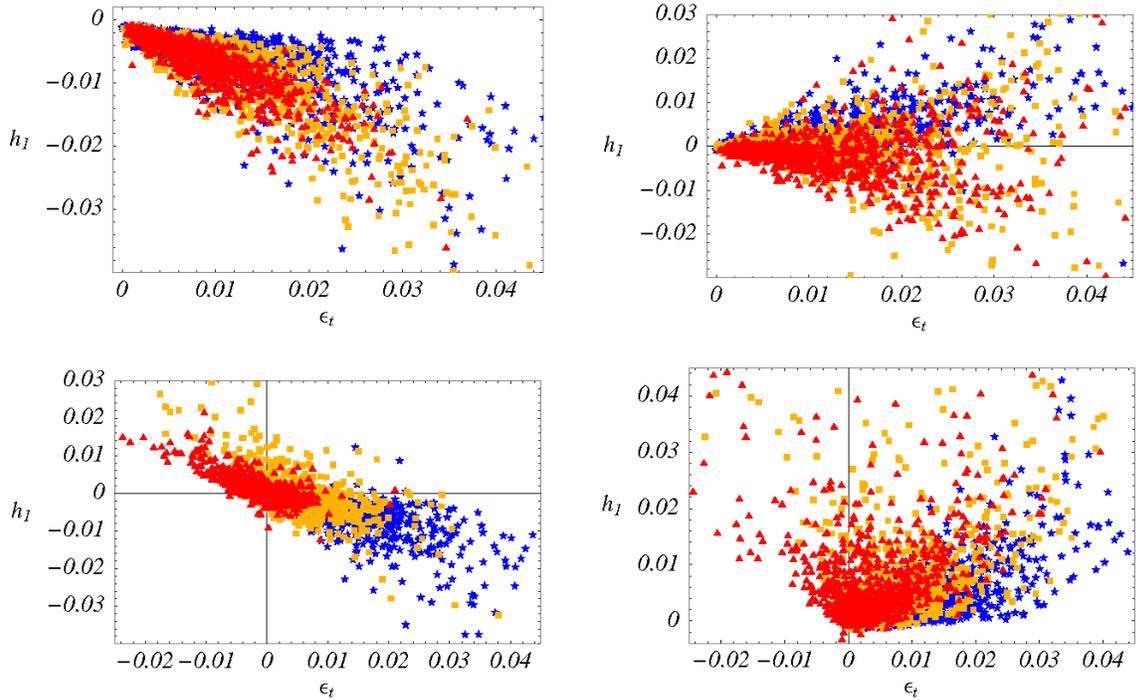}
  \vspace*{-0.2cm}
  \caption{The correction $h_1$ versus $\epsilon_t$, for $M_A=200\,$GeV 
  and negative $\mu$. Plots in the first row have 
  $\{ {\rm sgn} (\widetilde{A}_t), {\rm sgn} (\widetilde{A}_b) \}
  =\{+1,+1\}$ and $\{-1,+1\}$, and 
  $\{ {\rm sgn} (\widetilde{A}_t), {\rm sgn} (\widetilde{A}_b)\}
  =\{+1,-1\},\,\{-1,-1\}$ in the second row. Black, lightgray and 
  darkgray points (blue, yellow and red in colored plots) correspond to 
  $\tan\beta=10,\,35$ and $60$, respectively.}
  \label{fig7}
  \end{center}
\end{figure}

The dependence of the Higgs sector correction $h_2$ versus $\epsilon_b$, 
for three representative values of $\tan\beta=10,35,60$, $M_A=200\,$GeV, 
and all other SUSY parameters scanned randomly in the above intervals 
is  shown Figure~\ref{fig6}.  The perturbativity cut on $y_b$ has a strong 
effect, since it eliminates points with large negative values $\epsilon_b$ 
given a value of $\tan\beta$.  The value of $\epsilon_b$ at which this 
happens can be estimated by $\epsilon_b \tan\beta\approx -0.5$, 
which gives $-\epsilon_b\approx 0.05,0.014,0.008$ for $\tan\beta=10,35,60$, 
respectively, in agreement with the figures. At these points we 
observe a rapid increase of $|h_2|$ driven by the term
\begin{equation}
- 3 \mu^3 \,y_b^4 \widetilde{A}_b K_2(m_{\tilde Q},m_{\tilde b_R})
\end{equation}
in (\ref{eq:lambda7}), which gives a contribution to $h_2 \sim 
-\delta\lambda_7$ with sign opposite to $\widetilde{A}_b$.
Point i/ discussed above also applies to the case of negative 
$\mu$ taking into the account the reversed signs of $\epsilon_{b,t}$.  
Finally, Figure~\ref{fig7} shows the the Higgs correction $h_1$ versus 
$\epsilon_t$ for $\tan\beta=10,35,60$. Compared to the case of 
positive $\mu$, the sign of $\widetilde{A}_b$ has now a relevant effect, 
especially for larger values of $\tan\beta$, due to the contribution
of the $\widetilde{A}_b$ term in $z_{12}$. However, the growth with 
$y_b$ is less pronounced than in case of $h_2$, since $h_1$ does not 
contain terms proportional to $y_b^4$. 

\section{Summary}

This paper has been motivated by previous work \cite{Carena:1999py} 
that systematically investigated the resummation of  
SUSY QCD large-$\tan\beta$ effects in the decoupling limit, where the 
standard model particles and Higgs scalars remain light, but which did not 
consider electroweak effects. We performed a complete one-loop matching of 
the MSSM to a two-Higgs doublet model for all couplings that 
are absent or suppressed in the MSSM with large $\tan\beta$, 
keeping all contributions that are $\tan\beta$-enhanced relative to the 
suppressed tree terms. This includes complete expressions for the 
kinetic mixing and Higgs self couplings $\delta\lambda_{5-7}$, which 
have not been given previously. 

Our result confirms, as expected, that a resummation of large-$\tan\beta$ 
effects to arbitrary loop order is necessary only 
for the bottom and tau Yukawa couplings, if the bottom and 
tau mass are used as parameters of the MSSM. It clarifies a point 
left open in \cite{Carena:1999py}, namely how the renormalization 
of $\tan\beta$ affects this resummation. We find that the standard 
expression for the resummed Yukawa coupling is valid in renormalization 
schemes where the counterterms (shifts) of $\tan\beta$ and the vacuum 
expectation value $v_d$ do not receive  large-$\tan\beta$ contributions. 
In other schemes, 
a finite counterterm must be explicitly included in the resummation 
formula.

Besides the known ``wrong-Higgs'' Yukawa couplings, all other Higgs-sector 
effects that feed into the Yukawa couplings by modifying the definition 
of the physical Higgs fields at one loop can be parameterized 
in terms of two couplings $h_1$, $h_2$ dominated by kinetic mixing 
and the Higgs self-coupling $\delta\lambda_7$, respectively. 
The same $h_2$ 
also enters the effective trilinear Higgs couplings. Our numerical 
study suggests that, where present, the Higgs-sector effect $h_2$ 
is more important than the ``wrong-Higgs'' Yukawa couplings, and 
we identified the dominant SUSY parameter dependences. Our result 
extends and corrects a previous result \cite{Freitas:2007dp} 
that included the kinetic-mixing effect, but not the one from the 
one-loop induced Higgs couplings.

Although obtained in the decoupling limit, the effective one-loop 
couplings derived in this paper should be useful to obtain simple 
analytic estimates of the leading quantum corrections to those Yukawa and 
Higgs trilinear interactions, that are suppressed at tree-level, 
such as the $\bar b b h^0 $ coupling.

\vspace*{1em}

\noindent
\subsubsection*{Acknowledgement}

We thank U.~Nierste and S.~J\"ager for useful discussions. 
This work is supported in part by the 
DFG Sonder\-forschungsbereich/Transregio~9 
``Computergest\"utzte Theoretische Teilchenphysik'' 
and the Swiss National Science Foundation (SNF). 
M.B. acknowledges hospitality from the University of Z\"urich and the 
CERN theory group, where part of this work was performed.
Feynman diagrams have been drawn with the packages 
{\sc Axodraw}~\cite{Vermaseren:1994je} and 
{\sc Jaxo\-draw}~\cite{Binosi:2003yf}.


\end{document}